%% file: paper.tex
\documentclass[12pt]{article}
\usepackage{amsmath,cite,epsfig,a4,times,euler,euscript}
\renewcommand{\baselinestretch}{1.1}
\newcommand{\myTitle}[1]{\begin{center}{\bf\Huge #1}\\[5ex]\end{center}}
\newcommand{\myAuthor}[1]{\begin{center}{\Large #1}\\[2ex]\end{center}}
\newcommand{\myAffiliation}[1]{\\[1ex]{\it\large #1}}
\newcommand{\myEmail}[1]{}
\newcommand{\myDate}{\begin{center}{\large\today}\\[5ex]\end{center}}
\newcommand{\myAbstract}[1]{\begin{center}\renewcommand{\baselinestretch}{1}{\bf Abstract}\\[2ex]\parbox{0.8\linewidth}{\small\hspace{15pt} #1}\end{center}\vspace{\baselineskip}}
\newcommand{\myReport}[1]{\hspace{\fill} #1}
\newcommand{\myPreprint}[1]{}
\newcommand{\myKeywords}[1]{}
\newcommand{\myFigure}[1]{\begin{figure}\begin{center}#1\end{center}\end{figure}}
\newcommand{\myTable}[1]{\begin{table}\begin{center}#1\end{center}\end{table}}
\newcommand{\myScript}[1]{\EuScript{#1}}
\newcommand{\fudgea}{\\[0.3ex]}
\newcommand{\fudgeb}{\\[-0.6ex]}
\newcommand{\Appendix}[1]{Appendix~\ref{#1}}   
\newcommand{\Section}[1]{Section~\ref{#1}}
\newcommand{\Table}[1]{Table~\ref{#1}}
\newcommand{\Figure}[1]{Fig.\ref{#1}}
\newcommand{\Equation}[1]{Eq.(\ref{#1})}
\newcommand{\ie}{{\it i.e.}}
\newcommand{\mult}{n}
\newcommand{\rank}{r}
\newcommand{\Dim}{D}
\newcommand{\gs}{\mathrm{g}_{\mathrm{s}}}

\newcommand{\Aa}{A}
\newcommand{\Tr}{\mathrm{Tr}}
\newcommand{\SUgen}{T}
\newcommand{\ixm}{i}
\newcommand{\ixn}{j}
\newcommand{\ixk}{k}
\newcommand{\ixl}{l}
\newcommand{\spi}{x}
\newcommand{\spj}{y}
\newcommand{\spk}{z}
\newcommand{\lom}{\mu}
\newcommand{\lon}{\nu}
\renewcommand{\lor}{\rho}
\newcommand{\los}{\sigma}
\newcommand{\lol}{\lambda}
\newcommand{\VV}{V}
\newcommand{\VVR}{\bar{V}}
\newcommand{\VW}{W}
\newcommand{\VWR}{\bar{W}}
\newcommand{\PPR}{\bar{P}}
\newcommand{\RR}{R}
\newcommand{\Rtwo}{\myScript{R}_{2}}
\newcommand{\VX}{X}
\newcommand{\VY}{Y}
\newcommand{\mom}{p}
\newcommand{\mass}{m^2}
\newcommand{\metric}{g}
\newcommand{\GG}{G}
\newcommand{\Gb}{\bar{G}}
\newcommand{\GT}{\myScript{G}}
\newcommand{\UT}{\myScript{U}}
\newcommand{\FF}{F}
\newcommand{\FT}{\myScript{F}}
\newcommand{\qq}{q}
\newcommand{\rr}{r}
\newcommand{\dd}{\myScript{D}}
\newcommand{\Pol}{\varepsilon}
\newcommand{\ipol}{i}
\newcommand{\HH}{H}
\newcommand{\Hb}{\bar{H}}
\newcommand{\UU}{U}
\newcommand{\HT}{\myScript{H}}

\newcommand{\imag}{\mathrm{i}}
\newcommand{\graph}[3]{\raisebox{-#3pt}{\epsfig{file=#1.eps,width=#2pt}}}
\newcommand{\TT}{\myScript{T}}
\newcommand{\BB}{B}
\newcommand{\Amp}{\myScript{A}}

\newcommand{\ncol}{N_{\mathrm{c}}}
\newcommand{\nfer}{N_{\mathrm{f}}}
\renewcommand{\Re}{\mathrm{Re}}
\renewcommand{\Im}{\mathrm{Im}}

\newcommand{\Ord}{\mathcal{O}}
\newcommand{\GZ}{\cite{Giele:2008bc}}

\begin{document}

\myReport{IFJPAN-IV-2009-5}
\myPreprint{IFJPAN-IV-2009-5\\arXiv:0905.1005 [hep-ph]}

\myTitle{%
Multi-gluon one-loop amplitudes\fudgea
using tensor integrals%
\footnote{%
This work was partially supported by RTN European Programme, MRTN-CT-2006-035505 (HEPTOOLS, Tools and Precision Calculations for Physics Discoveries at  Colliders)
}
}

\myAuthor{%
A.\ van Hameren%
\myAffiliation{%
The H.\ Niewodnicza\'nski Institute of Nuclear Physics\fudgeb
Polisch Academy of Sciences\\
Radzikowskiego 152, 31-342 Cracow, Poland%
\myEmail{hameren@ifj.edu.pl}
}
}

\myDate

\myAbstract{%
An efficient numerical algorithm to evaluate one-loop amplitudes
using tensor integrals is presented. In particular, it is shown
by explicit calculations that for ordered QCD amplitudes with a
number of external legs up to 10, its performance is competitive
with other methods.
}

\myKeywords{NLO computations, QCD}

%

\section{Introduction\label{Sec:intro}}
\input{introduction.tex}

\section{Tensor reduction\label{Sec:reduction}}
\input{tensors.tex}

\section{Ordered gluon one-loop amplitudes\label{Sec:recursion}}
In the following, we wil derive recursive relations for the tensors to be contracted with the tensor integrals in order to arrive at one-loop amplitudes.
First we repeat the known tree-level relations to introduce some notation.

\subsection{Recursive relations for ordered gluon tree-level amplitudes}
\input{treelevel.tex}

\subsection{Recursive relations for ordered gluon one-loop amplitudes}
\input{looplevel.tex}

\subsection{Including ghost and quark loops}
\input{ghostandr2.tex}

\section{Results\label{Sec:results}}
\input{results.tex}

\section{Conclusion\label{Sec:conclusion}}
\input{conclusion.tex}

\subsection*{Acknowledgments}
The author would like to thank C.~G.~Papadopoulos for useful discussions and comments.


\input{bibliography.tex}

\begin{appendix}
\section{Reproduction of existing results\label{App1}}
\input{appendixA.tex}

\section{Quark loops\label{App2}}
\input{appendixB.tex}

\end{appendix}
%
%
\end{document}

%% file: introduction.tex
In order to deal with the data from the experiments at LHC for the study of elementary particles, signals and potential backgrounds for new physics have to be under control at sufficient accuracy~\cite{Bern:2008ef}.
In particular, hard processes with high multiplicities, involving many particles or partons, cannot be neglected.
On top of that, such processes have to be dealt with at the next-to-leading order (NLO) level to, for example, reduce the scale dependence of observables and to have a better description of the shape of their distributions.

An important part of a NLO calculation concerns the one-loop amplitude.
Recently, impressive results have been published for one-loop QCD amplitudes for very high numbers of partons \cite{Giele:2008bc,Lazopoulos:2008ex,Winter:2009kd}.
They were obtained with the so-called unitarity-approach.
Originally restricted to analytical calculations \cite{Bern:1994cg,Bern:1994zx,Bern:1993mq,Bern:1997sc,Bern:1994fz}, the potential of this method in a numerical approach became, after the crucial input from \cite{Britto:2004nc}, clear with the work of \cite{Ossola:2006us,Ossola:2007bb,Ossola:2008xq} and \cite{Giele:2008ve}.
It is considered an alternative to the ``traditional'' approach involving tensor integrals.
Both approaches expand the one-loop amplitude in terms of a basis set of one-loop functions.
In the unitarity-approach, this set consists of scalar-integrals up to $4$-point or $5$-point functions, and it aims at determining the coefficients directly.
In the ``tensor-approach'', the basis set is larger and consists of tensor integrals or their coefficients functions when expanded in terms of Lorentz-covariant objects~%
\cite{
Passarino:1978jh
,vanOldenborgh:1989wn
,Ezawa:1990dh
,Belanger:2003sd
,Davydychev:1991va
,Tarasov:1996br
,Fleischer:1999hq
,Bern:1992em
,Bern:1993kr
,Binoth:1999sp
,Duplancic:2003tv
,Giele:2004iy
,Giele:2004ub
,Ellis:2005zh
,Binoth:2005ff
,delAguila:2004nf
,vanHameren:2005ed
,Denner:2002ii
,Denner:2005nn
,Diakonidis:2008dt
,Diakonidis:2008ij
}.
Also these basis-functions are eventually calculated by expressing them in terms of a smaller set of scalar-integrals, but this happens in a, for the particular method, universal way, independent of the amplitude.

Multiplicities with up to $20$ partons as achieved in \cite{Giele:2008bc,Lazopoulos:2008ex} are unattainable in the tensor-approach because of the asymptotic computational complexity of the latter.
It arises because the basis set contains $n$-point functions where $n$ goes up to the total number of external legs of the amplitude.
Let us make a crude comparison between the unitarity-approach and the tensor-approach of \cite{delAguila:2004nf} in which the basis set consists of ``normal'' tensor integrals carrying explicit Lorentz-indices.
A first step in the analysis of the computational complexity of the two methods is the determination of the number of coefficients to be evaluated in case ordered amplitudes have to be calculated.
For the unitarity-approach which determines coefficients up to $4$-point functions, it is given by
%
\begin{equation}
\binom{n}{4} + \binom{n}{3} + \binom{n}{2} + \binom{n}{1}
=
\frac{14n + 11n^2 - 2n^3 + n^4}{24}
~.
\end{equation}
%
The number of tensor integrals is, using the fact that only symmetric tensors have to be considered of a rank not higher than the multiplicity,
%
\begin{equation}
\sum_{k=1}^{n}\sum_{l=0}^{k}\binom{n}{k}\binom{l+3}{l}
=
\frac{-384 + 592 n + 203 n^2  + 26 n^3  + n^4}{384} \, 2^n
~.
\label{Eq153}
\end{equation}
%
Here, all tensors up to the maximal ranks have been included.
The number is obviously much larger than the number of scalar functions.
The asymptotic behavior of $2^n$ for the tensor integrals is particularly disastrous.
It is a result of the expansion in terms of $n$-point integrals.
The accompanying $n^4$-behavior stems from the symmetric tensor components.
Of course this is not the whole story.
Also the operations to be performed in order to determine the coefficients have to be taken into account.
For the unitarity-approach as presented in \cite{Giele:2008bc} for example, this leads to a final computational complexity of $\Ord(n^9)$.
In this write-up, we will see that, by using recursion on both the tensor integrals and their coefficients, the complexity as given in \Equation{Eq153} does not change.
Although the asymptotic complexity is exponential as opposed to the polynomial complexity of the unitarity-approach, it may be competitive for moderate values of $n$, and in fact, we will see that in practice it is up to $n=10$.

Besides the computational complexity, also the numerical stability is an important issue concerning one-loop calculations.
It is typically related to inverse Gram-determinants approaching zero.
When using tensor integrals, this issue is isolated to the calculation of the tensor integrals themselves.
In that sense, it allows for universal solutions, and several methods to achieve this exist.
As a last resort, the calculation of the tensor integrals can be performed at higher precision level for phase-space points at which numerical instabilities occur, and the decision to do so can be made at relatively low cost.
In the unitarity-approach, numerical instabilities can show up in the computation of the coefficients for the scalar functions.
At the moment, there is no better cure known than to increase the precision level for the full calculation for phase-space points at which numerical instabilities occur.
Fortunately, also in this case the decision is relatively cheap.

A final issue worth mentioning is the potential towards automation of the method.
The unitarity-approach proofs to be successful in this respect because it optimally allows for the use of existing tree-level machinery related to the calculation of off-shell currents or ``sub-amplitudes''.
We will see that the method presented here allows for the same.
In particular, it is completely numerical and no computer algebra is involved.

The outline of the paper is as follows.
In \Section{Sec:reduction} the tensor reduction is addressed, and in \Section{Sec384} tensor symmetrization, which is crucial for the efficiency of the presented algorithm.
Recursive relations for one-loop amplitudes are presented in \Section{Sec:recursion}, and in \Section{Sec:results} results can be found obtained with the help of an explicit implementation of the algorithms in the foregoing sections.
The conclusions in \Section{Sec:conclusion} finally close the paper.
%

%% file: tensors.tex
Tensor integrals are usually calculated using recursive equations relating high-multiplicity and high-rank tensor integrals to lower-multiplicity and lower-rank ones.
Tensor reduction is this formal process,
in practice the opposite process, tensor building, is performed.
The multiplicity $\mult$ and and rank $\rank$ are defined with the formula
%
\begin{equation}
\TT_{\mult,\rr}^{\lon_{1}\lon_{2}\cdots\lon_{\rr}}
=
\int\frac{d^\Dim\qq}{\imag\pi^{\Dim/2}}
\frac{\qq^{\lon_{1}}_{4}\qq^{\lon_{2}}_{4}\cdots\qq^{\lon_{\rr}}_{4}}
     {\prod_{j=1}^{\mult}[(\qq+\mom_{j})^2-\mass_{j}]}
~.
\label{Eq225}
\end{equation}
%
As the formula suggests, we consider tensor integrals defined in $\Dim$ dimensions, but only with $4$-dimensional components of the integration momentum in the numerator.
This will lead to a calculation of the one-loop amplitude within the scheme of \cite{Weinzierl:1999xb}, which asks for a finite counterterm in order to arrive at gauge-invariant results and to cast the result into other schemes like 't~Hooft-Veltman or FDH.
This finite counterterm is exactly given by the so-called $\Rtwo$-term, showing up explicitly in the OPP unitarity-approach as part of the rational terms \cite{Ossola:2008xq}, and which is shown how to be determined in \cite{Draggiotis:2009yb}.


The asymptotic computational complexity given in \Equation{Eq153} does not increase when the operations needed to calculate the tensor integrals are taken into account, because for high-$n$ each integral can be obtained using a fixed number of lower integrals, independent of $n$ or $r$.
This can be easily understood as follows.
Using the fact that we can write
%
\begin{equation}
2(\mom_{j}-\mom_{\mult})\cdot\qq
=
[(\qq+\mom_{j})^2-\mass_{j}]
-[(\qq+\mom_{\mult})^2-\mass_{\mult}]
+\mass_{j}-\mom_{j}^2-\mass_{\mult}+\mom_{\mult}^2
~,
\end{equation}
%
we have
%
\begin{multline}
2(\mom_{j}-\mom_{\mult})_{\lon_{\rr}}
\TT_{\mult,\rr}^{\lon_{1}\lon_{2}\cdots\lon_{\rr}}
=
 \TT_{\mult-1,\rr-1}^{\lon_{1}\lon_{2}\cdots\lon_{\rr-1}}(j)
-\TT_{\mult-1,\rr-1}^{\lon_{1}\lon_{2}\cdots\lon_{\rr-1}}(\mult)
\\
+(\mass_{j}-\mom_{j}^2-\mass_{\mult}+\mom_{\mult}^2)
\,\TT_{\mult,\rr-1}^{\lon_{1}\lon_{2}\cdots\lon_{\rr-1}}
~,
\end{multline}
%
where $\TT_{\mult-1,\rr-1}^{\lon_{1}\lon_{2}\cdots\lon_{\rr-1}}(j)$ is obtained from $\TT_{\mult,\rr-1}^{\lon_{1}\lon_{2}\cdots\lon_{\rr-1}}$ by removing the $j$-th denominator.
Choosing $4$ different vectors $\mom_{j}$ appearing in the denominators, we get $4$ relations, enough to determine the $4$ integrals $\TT_{\mult,\rr}^{\lon_{1}\lon_{2}\cdots\lon_{\rr}}$ with the first $\rr-1$ Lorentz indices fixed.
So $4$ tensor integrals can be determined using $12$ lower integrals. 
Although very straightforward, this is numerically not necessarily the best method to calculate tensor integrals, since it involves the inversion of a $4\times4$-matrix.
The method presented in \cite{delAguila:2004nf} involves the square-root of an inverse Gram determinant of only $3$ vectors.
Also these can be chosen out of $\mult-1$ denominators, and for high $\mult$ the probability that a phase-space point is such that all combinations lead to small Gram determinants is rather low.

For low-$\mult$ integrals, \ie\ for $\mult\leq4$, the previous statement is obviously not true, but several recipes and their implementations to deal with numerical instabilities exist.
Notice that, in renormalizable gauges, $\rank\leq\mult$, so that for low $\mult$ also the cost, for example, of converting Passarino-Veltman functions calculated following \cite{Denner:2005nn} to tensor integrals like above is acceptable.
In fact, to obtain the results in this write-up, the ``Alternative Passarino-Veltman-like reduction'' from \cite{Denner:2005nn} was used for the $4$-point integrals.
It can easily be predicted when this method fails, in which case the method from \cite{delAguila:2004nf} was used.
For the $3$-point functions, conventional Passarino-Veltman reduction was used.

The end-points of the tensor reduction are scalar integrals.
Also these can recursively be reduced further, and eventually be expressed in terms of $4$-point scalar functions.
For the application in this write-up, the unitarity-approach as presented in \cite{Ossola:2006us} was used to express scalar functions into $4$-point functions directly.
This choice, of course, is not in correspondence with the ``recursivity philosophy'', and in fact it strictly speaking increases the asymptotic computational complexity to $\Ord(\mult^52^\mult)$.%
\footnote{Roughly speaking $\Ord(2^\mult)$ scalar functions need $\Ord(\mult^4)$ coefficients for the $4$-point functions, each of which involves the evaluation of $\Ord(\mult)$ denominators.}
In practice, however, it appears to be rather numerically stable and efficient, in particular because the formulations of \cite{delAguila:2004nf} and \cite{Ossola:2006us} are compatible to large extend, avoiding the re-calculation of some overhead.
Furthermore, for the scalar-functions, no numerator functions have to be evaluated, and only the coefficients for the $4$-point functions have to be calculated, avoiding the computationally more challenging issues coming with the method of \cite{Ossola:2006us} in general.

The scalar one-loop $1$-point, $2$-point, $3$-point and $4$-point functions, finally, were evaluated with {\tt OneLOop}~\cite{vanHameren:2009dr}.
Also the tensor $2$-point functions were evaluated with this program.

\section{Tensor symmetrization\label{Sec384}}
%
Also for high-rank tensors, one has to deal with tensor-contractions in the end.
Contracting rank-$\rank$ tensors with $4^\rank$ tensor components seems hopeless, and the solution to this problem is tensor symmetrization.
Here, we use the fact that the tensor integrals \Equation{Eq225} are symmetric, and a tensor integral of rank $\rank$ has only
%
\begin{equation}
\binom{\rank+3}{\rank}
=
\frac{6 + 11\rr + 6\rr^2 + \rr^3}{6}
\end{equation}
%
independent components.
So for symmetric tensors, the issue of contraction does not look hopeless at all.
We only have to make sure we can calculate the symmetrized coefficients to be contracted with the tensor integrals directly.
By {\em symmetrization\/} we mean adding tensor-components which are multiplied by the same tensor integral together, so
%
\begin{eqnarray}
&&T_{\rr=2}^{\{1,2\}}
=
T_{\rr=2}^{1,2} + T_{\rr=2}^{2,1}
\quad,\quad
T_{\rr=3}^{\{1,2,2\}}
=
T_{\rr=3}^{1,2,2} + T_{\rr=3}^{2,1,2} + T_{\rr=3}^{2,2,1}
\nonumber\\
&&T_{\rr=3}^{\{1,2,3\}}
=
 T_{\rr=3}^{1,2,3} + T_{\rr=3}^{2,3,1} + T_{\rr=3}^{3,1,2}
+T_{\rr=3}^{3,2,1} + T_{\rr=3}^{2,1,3} + T_{\rr=3}^{1,3,2}
~,
\end{eqnarray}
%
etc..
In general, a symmetric tensor $T_{\rr}^{\{\lon_1\lon_2\cdots\lon_{\rr}\}}$ of rank $\rr$ with $4$-dimensional indices can be represented as
%
\begin{equation}
T_{\rr}^{\{\lon_1\lon_2\cdots\lon_{\rr}\}}
=
S^{\rr}_{n_0,n_1,n_2,n_3}
\end{equation}
%
where $n_{\lom}$ is the number of indices referring to dimension $\lom$.
These numbers satisfy $n_0+n_1+n_2+n_3=\rr$.
Now suppose we have a linear recursive relation between tensors of the type
%
\begin{equation}
T_{\rr}^{\lon_1\lon_2\cdots\lon_{\rr}}
=
T_{\rr-1}^{\lon_1\lon_2\cdots\lon_{\rr-1}}K_{\rr}^{\lon_{\rr}}
\label{Eq371}
\end{equation}
%
with $T_{1}^{\lon}=K_{1}^{\lon}$.
The solution is a product of the components of the vectors $K_{1}$ to $K_{\rank}$.
To calculate the symmetrized product, we can cast the relation in the form
%
\begin{eqnarray}
S^{\rr}_{n_0,n_1,n_2,n_3}
&=&
 S^{\rr-1}_{n_0-1,n_1,n_2,n_3}\,K_{\rr}^0 
+S^{\rr-1}_{n_0,n_1-1,n_2,n_3}\,K_{\rr}^1
\nonumber\\ 
&+&S^{\rr-1}_{n_0,n_1,n_2-1,n_3}\,K_{\rr}^2 
+S^{\rr-1}_{n_0,n_1,n_2,n_3-1}\,K_{\rr}^3
~,
\end{eqnarray}
%
with the convention that $S^{\rr}_{n_0,n_1,n_2,n_3}$ is identically zero whenever any of the indices is negative.
%


The relation expressed by \Equation{Eq371} seems rather trivial, but we will see that the tensor components we have to calculate satisfy very similar relations.
The main difference will be that the simple multiplications on the r.h.s.\ will be replaced by more complicated contractions.
This does not have any influence on the possibility to calculate symmetrized components directly, nor on the asymptotic computational complexity.
In fact, like the calculation of the tensor integrals, also the calculation of the tensor components does not increase the asymptotic complexity given in \Equation{Eq153}.
This stems from the facts that the number of operations to be performed to calculate a tensor given the lower tensors is constant, and that the number of tensors entering the recursive equation is equal to the number of tensor integrals, \ie, no intermediate ``auxiliary'' tensors have to be calculated.
%
The equivalents of the vectors $K_{\rr}^{\lom}$ above will essentially consist of tree-level off-shell currents, which are computed at a cost of $\Ord(\mult^4)$.
%

%

%% file: treelevel.tex
The recursive relations for tree-level gluon off-shell currents are given by \cite{Berends:1987me}
%
\begin{eqnarray}
\Aa^{\lom}_{\ixm,\ixn}
=
\frac{-\imag}{\mom_{\ixm,\ixn}^2}
\left[
\sum_{\ixk=\ixm}^{\ixn-1}
\VV^{\lom}_{\lon\lor}(\mom_{\ixm,\ixk},\mom_{\ixk+1,\ixn})
\Aa^{\lon}_{\ixm,\ixk}
\Aa^{\lor}_{\ixk+1,\ixn}
+
\sum_{\ixk=\ixm}^{\ixn-2}
\sum_{\ixl=\ixk+1}^{\ixn-1}
\VW^{\lom}_{\lon\lor\los}
\Aa^{\lon}_{\ixm,\ixk}
\Aa^{\lor}_{\ixk+1,\ixl}
\Aa^{\los}_{\ixl+1,\ixn}
\right]
\label{Eq137}
\end{eqnarray}
%
with
%
\begin{equation}
\VV^{\lom}_{\lon\lor}(\mom_{1},\mom_{2})
=
\frac{\imag\gs}{\sqrt{2}}
[\; (\mom_{1}-\mom_{2})^{\lom}\metric_{\lon\lor}
   +(\mom_{1}+2\mom_{2})_{\lon}\,\metric_{\lor}^{\lom}
   -(\mom_{2}+2\mom_{1})_{\lor}\,\metric_{\lon}^{\lom}\;]
\end{equation}
%
and
%
\begin{equation}
\VW^{\lom}_{\lon\lor\los}
=
\frac{\imag\gs^2}{2}
[\;2\metric^{\lom}_{\lor}\,\metric_{\lon\los}
   -\metric^{\lom}_{\lon}\,\metric_{\lor\los}
   -\metric^{\lom}_{\los}\,\metric_{\lor\lon}\;]
~.
\end{equation}
%
The starting points $\Aa^{\lom}_{\ixm,\ixm}=\Pol^{\lom}_{\ixm}$ of these recursive equations are the polarizations vectors of the external gluons, and we denote
%
\begin{equation}
\mom_{\ixm,\ixn}
=
\sum_{\ixk=\ixm}^{\ixn}\mom_{\ixk}
\label{Eq494}
\end{equation}
%
where $\mom_{\ixk}$ is the momentum of gluon $\ixk$.
For $\ixm<\ixn$ we define $\mom_{\ixm,\ixn}=0$.
If $\mom_{1,\mult}=0$, then
%
\begin{equation}
\Amp_{\mult}(1,2,\ldots,\mult)
=
\metric_{\lon\lom}\,\Pol^{\lon}_{\mult}\,\mom_{\mult}^2\Aa^{\lom}_{1,\mult-1}
\label{Eq193}
\end{equation}
%
is the tree-level color-ordered amplitude for gluon $1$ to $\mult$.
The full tree-level amplitude for the $\mult$ gluons is then given by~\cite{Mangano:1987xk}
%
\begin{equation}
\myScript{M}_{\mult}
=
\sum_{\pi\in{}S_{\mult}/Z_{\mult}}
\Tr(\SUgen^{a_{\pi(1)}}\SUgen^{a_{\pi(2)}}\cdots\SUgen^{a_{\pi(\mult)}})
\,\Amp_{\mult}(\pi(1),\pi(2),\ldots,\pi(\mult))
~,
\label{Eq210}
\end{equation}
%
where $a_1,a_2,\ldots,a_{\mult}$ are the color indices of the gluons and $\SUgen^a$ are the generators of $\mathrm{SU}(\ncol)$.
The sum is over all permutations of the gluons except the cyclic permutations.
The off-shell currents satisfy
$\metric_{\lon\lom}\mom_{\ixm,\ixn}^{\lon}\Aa^{\lom}_{\ixm,\ixn}=0$%
, and the three-point vertex can be reduced to
%
\begin{equation}
\VV^{\lom}_{\lon\lor}(\mom_{1},\mom_{2})
=
\frac{\imag\gs}{\sqrt{2}}
[\; (\mom_{1}-\mom_{2})^{\lom}\metric_{\lon\lor}
   +2\mom_{2\lon}\,\metric_{\lor}^{\lom}
   -2\mom_{1\lor}\,\metric_{\lon}^{\lom}\;]
~.
\end{equation}
%
The recursive equation may be represented diagrammatically by
%
\begin{equation}
\graph{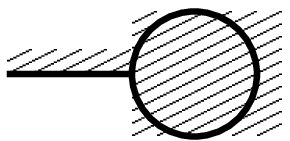}{36}{6}
=
\graph{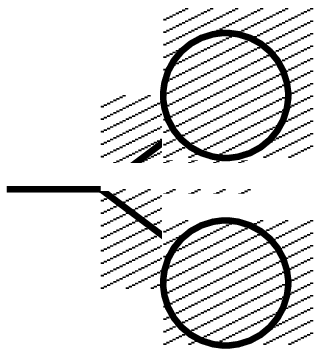}{41}{20}
+
\graph{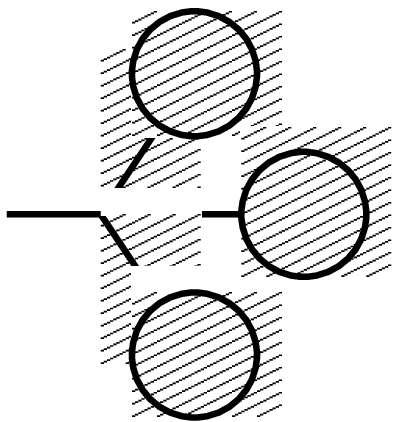}{51}{26}
\quad.
\label{Eq209}
\end{equation}
%
Even for a diagrammatic representation this formula is rather rudimentary, but it encodes enough information for our purpose.
For a more detailed description of the recursive relation, we prefer to refer to \Equation{Eq137} instead of dressing up the diagrammatic representation.
%

%% file: looplevel.tex
A so called {\em color decomposition\/} as in \Equation{Eq210} also exists for one-loop amplitudes~\cite{Bern:1990ux}, and is given by
%
\begin{eqnarray}
\myScript{M}_{\mult}^{(1)}
&=&
\sum_{\pi\in{}S_{\mult}/Z_{\mult}}
\Tr(\SUgen^{a_{\pi(1)}}\cdots\SUgen^{a_{\pi(\mult)}})
\,\Amp^{(1)}_{\mult}(\pi(1),\ldots,\pi(\mult))
\label{Eq273}\\
&+&
\sum_{c=2}^{\lfloor\mult/2\rfloor+1}\sum_{\pi\in{}S_{\mult}/S_{\mult;c}}
\Tr(\SUgen^{a_{\pi(1)}}\cdots\SUgen^{a_{\pi(c-1)}})
\,\Tr(\SUgen^{a_{\pi(c)}}\cdots\SUgen^{a_{\pi(\mult)}})
\,\Amp^{(c)}_{\mult}(\pi(1),\ldots,\pi(\mult))
~,
\nonumber
\end{eqnarray}
%
where $S_{\mult;c}$ is the subset of $S_{\mult}$ that leaves the corresponding double trace structure invariant.
The objects $\Amp^{(1)}_{\mult}$ are called {\em primitive amplitudes\/}.
They only receive contributions from diagrams with a particular ordering of the gluons.
The partial amplitudes $\Amp^{(c)}_{\mult}$ can be calculated as linear combinations of permutations of the primitive amplitudes.

Given the definition of the primitive amplitudes, one can write down a recursive relation for off-shell currents from which the primitive amplitudes can be constructed following a relation like \Equation{Eq193}.
The blobs in the diagrammatic equation \Equation{Eq209} represent off-shell currents consisting of sums of tree-level diagrams.
We represent off-shell currents consisting of diagrams containing exactly $1$ loop by a blob with a hole, and we have
%
\begin{multline}
\graph{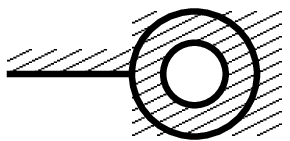}{36}{6}
=
  \graph{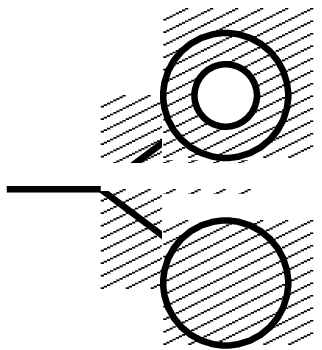}{41}{20}
+ \graph{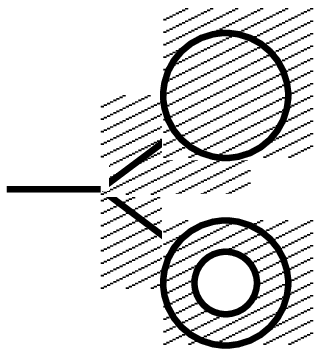}{41}{20}
+ \graph{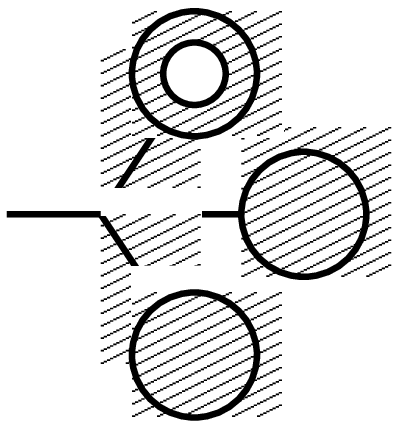}{51}{26}
+ \graph{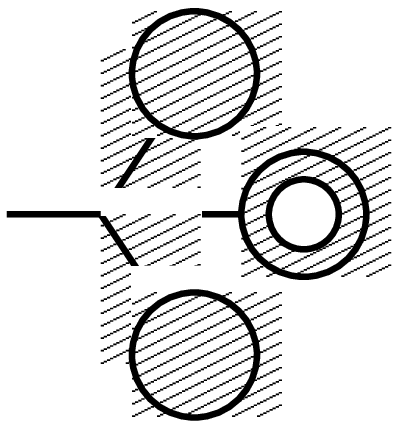}{51}{26}
+ \graph{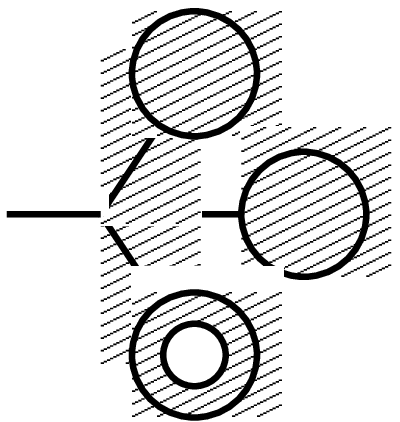}{51}{26}
\\
+ \graph{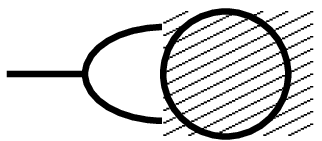}{41}{6}
+ \graph{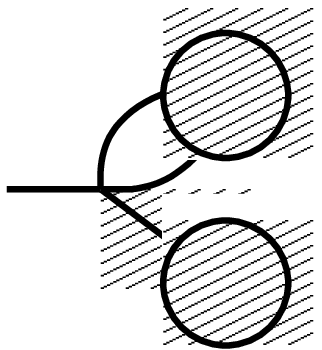}{41}{20}
+ \graph{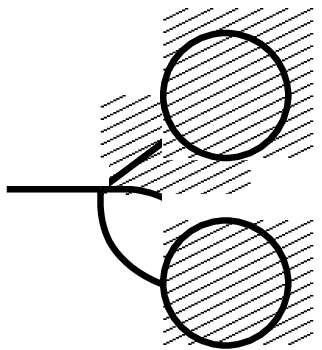}{41}{20}
\quad.
\label{Eq240}
\end{multline}
%
Concerning the first line, it is clear that, since the result may only consist of one-loop diagrams, exactly one blob with a hole must be connected to a vertex.
Since we are considering {\em ordered\/} amplitudes, all distributions of the blob with a hole over the different legs of a vertex have to be represented separately.

The actual loops are generated in the second line of \Equation{Eq240}.
Both the second and the third term on this line have to be added explicitly because of the ordering.
These loops are constructed from tree-level off-shell currents with one auxiliary gluon with momentum and polarization vector, say, $\qq$ and $\Pol^{\lom}(\ipol)$ respectively.
The index $\ipol$ runs from $1$ to $4$ such that
%
\begin{equation}
\sum_{\ipol=1}^4\Pol^{\lon}(\ipol)\Pol^{\lor}(\ipol)=\metric^{\lon\lor}
~.
\end{equation}
%
This gluon is supposed to be virtual, and thus off-shell.
The polarization vector has no real physical meaning, and just plays the r\^ole of the end-point of the gluonic line.
We will now introduce objects $\GG^{\lol\lom}_{\ixm,\ixn}(\qq)$ including this auxiliary gluon through the formal relations
\begin{eqnarray}
\graph{graph04}{41}{6}
&\leftrightarrow&
\frac{-\imag}{\mom_{\ixm,\ixn}^2}
\int\frac{d^4\qq}{\imag\pi^{2}}
\,\VV^{\lom}_{\lon\lor}(-\qq-\mom_{1,\ixm-1},\qq+\mom_{1,\ixn})
\,\GG^{\lon\lor}_{\ixm,\ixn}(\qq)
\nonumber\\
\graph{graph05}{41}{20}
&\leftrightarrow&
\frac{-\imag}{\mom_{\ixm,\ixn}^2}
\sum_{\ixk=\ixm}^{\ixn-1}
\VW^{\lom}_{\lon\lor\los}
\,\Aa^{\los}_{\ixk+1,\ixn}
\int\frac{d^4\qq}{\imag\pi^{2}}
\,\GG^{\lon\lor}_{\ixm,\ixk}(\qq)
\label{Eq391}\\
\graph{graph06}{41}{20}
&\leftrightarrow&
\frac{-\imag}{\mom_{\ixm,\ixn}^2}
\sum_{\ixk=\ixm}^{\ixn-1}
\VW^{\lom}_{\lon\lor\los}
\,\Aa^{\lon}_{\ixm,\ixk}
\int\frac{d^4\qq}{\imag\pi^{2}}
\,\GG^{\lor\los}_{\ixk+1,\ixn}(\qq)
~.
\nonumber
\end{eqnarray}
%
The seemingly superfluous momentum shift $\mom_{1,\ixm-1}$ in the first line is to make sure that only inverse denominators of the form $(\qq+\mom_{1,k})^2$ appear in the calculation, and not for example $(\qq+\mom_{2,k})^2$.
This also means that the auxiliary external gluon in $\GG^{\lol\lom}_{\ixm,\ixn}(\qq)$ is carrying momentum $\qq+\mom_{1,\ixm-1}$ instead of $\qq$.
At this point, the question is how to assign a meaning to the relations above, and we will explain this in the following.

Since we are interested only in the contribution of {\em ordered\/} one-loop diagrams, the auxiliary gluon with momentum $\qq+\mom_{1,\ixm-1}$ must be the first one~%
\footnote{Or the last one, but we choose the first one.}
for every off-shell current $\GG^{\lol\lom}_{\ixm,\ixn}(\qq)$, so these off-shell currents satisfy
%
\begin{equation}
\GG^{\lol\lom}_{\ixm,\ixn}(\qq)
=
\graph{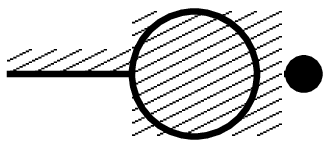}{45}{6}
=
  \graph{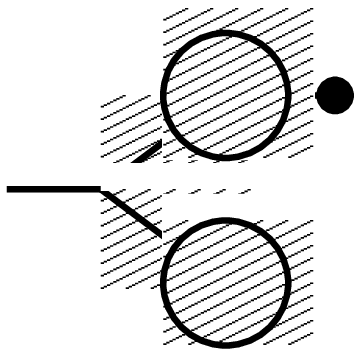}{50}{20}
+ \graph{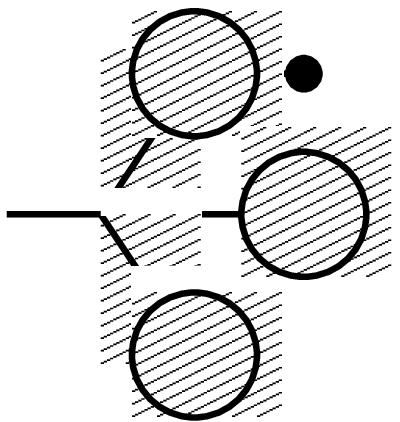}{50.5}{26}
\quad.
\label{Eq394}
\end{equation}
%
More explicitly, the relation is
%
\begin{multline}
\GG^{\lol\lom}_{\ixm,\ixn}(\qq)
=
\frac{-\imag}{(\qq+\mom_{1,\ixn})^2}
\Bigg[
\sum_{\ixk=\ixm-1}^{\ixn-1}
\VV^{\lom}_{\lon\lor}(\qq+\mom_{1,\ixk},\mom_{\ixk+1,\ixn})
\GG^{\lol\lon}_{\ixm,\ixk}(\qq)
\Aa^{\lor}_{\ixk+1,\ixn}
\\
+
\sum_{\ixk=\ixm-1}^{\ixn-2}
\sum_{\ixl=\ixk+1}^{\ixn-1}
\VW^{\lom}_{\lon\lor\los}
\GG^{\lol\lon}_{\ixm,\ixk}(\qq)\Aa^{\lor}_{\ixk+1,\ixl}\Aa^{\los}_{\ixl+1,\ixn}
\Bigg]
~.
\end{multline}
%
Notice that the sum over $\ixk$ starts with $\ixk=\ixm-1$: the case that $\GG^{\lol\lon}_{\ixm,\ixk}(\qq)$ does not contain any of the gluons $\ixm$ to $\ixn$ and for which it is given by
%
\begin{equation}
\GG^{\lol\lon}_{\ixm,\ixm-1}(\qq)
=
\metric^{\lol\lon}
\end{equation}
%
for every $\ixm$.
Introducing the symbol 
%
\begin{equation}
\VX^{\lom}_{\los\lon\lor}
=
\frac{\imag\gs}{\sqrt{2}}
[\; \metric^{\lom}_{\los}\metric_{\lon\lor}
   +\metric_{\lor}^{\lom}\metric_{\lon\los}
   -2\metric_{\lon}^{\lom}\metric_{\lor\los}\;]
\end{equation}
%
we can separate the $\qq$-dependent part of the $3$-point vertex and write
%
\begin{multline}
\GG^{\lol\lom}_{\ixm,\ixn}(\qq)
=
\frac{-\imag}{(\qq+\mom_{1,\ixn})^2}
\Bigg\{
\sum_{\ixk=\ixm-1}^{\ixn-1}
\left[
\VV^{\lom}_{\lon\lor}(\mom_{1,\ixk},\mom_{\ixk+1,\ixn})
+\VX^{\lom}_{\los\lon\lor}\qq^{\los}
\right]
\GG^{\lol\lon}_{\ixm,\ixk}(\qq)
\Aa^{\lor}_{\ixk+1,\ixn}
\\
+
\sum_{\ixk=\ixm-1}^{\ixn-2}
\sum_{\ixl=\ixk+1}^{\ixn-1}
\VW^{\lom}_{\lon\lor\los}
\GG^{\lol\lon}_{\ixm,\ixk}(\qq)\Aa^{\lor}_{\ixk+1,\ixl}\Aa^{\los}_{\ixl+1,\ixn}
\Bigg\}
~.
\label{Eq251}
\end{multline}
%
From these recursive equations, we can see that $\GG^{\lol\lom}_{\ixm,\ixn}(\qq)$ can be expressed as follows
%
\begin{equation}
\GG^{\lol\lom}_{\ixm,\ixn}(\qq)
=
\sum_{\dd\subset\{\ixm-1,\ixm,\ldots,\ixn\}}
\sum_{\rr=0}^{|\dd|-1}
\GT^{\lol\lom}_{\lon_1\lon_2\cdots\lon_{\rr}}(\dd)
\,\frac{\qq^{\lon_{1}}\qq^{\lon_{2}}\cdots\qq^{\lon_{\rr}}}
       {\prod_{j\in{}\dd}(\qq+\mom_{1,j})^2}
\label{Eq302}
\end{equation}
%
where $|\dd|$ is the number of elements in $\dd$, which is a subset of the set $\{\ixm-1,\ixm,\ldots,\ixn\}$ containing at least $\ixm-1$ and $\ixn$.
%
The tensors $\GT^{\lol\lom}_{\lon_1\lon_2\cdots\lon_{\rr}}(\dd)$ do not depend on $\qq$.
As an explicit example, we can write
%
\begin{multline}
\GG^{\lol\lom}_{2,4}(\qq)
=
  \frac{\GT^{\lol\lom}(1,4) + \GT^{\lol\lom}_{\lon}(1,4)\,\qq^{\lon}}{(\qq+\mom_{1,1})^2(\qq+\mom_{1,4})^2}
\\
+ \frac{  \GT^{\lol\lom}(1,2,4)
        + \GT^{\lol\lom}_{\lon}(1,2,4)\,\qq^{\lon}
        + \GT^{\lol\lom}_{\lon_1\lon_2}(1,2,4)\,\qq^{\lon_1}\qq^{\lon_2}}
       {(\qq+\mom_{1,1})^2(\qq+\mom_{1,2})^2(\qq+\mom_{1,4})^2}
\\
+ \frac{  \GT^{\lol\lom}(1,3,4)
        + \GT^{\lol\lom}_{\lon}(1,3,4)\,\qq^{\lon}
        + \GT^{\lol\lom}_{\lon_1\lon_2}(1,3,4)\,\qq^{\lon_1}\qq^{\lon_2}}
       {(\qq+\mom_{1,1})^2(\qq+\mom_{1,3})^2(\qq+\mom_{1,4})^2}
\\
+ \frac{  \GT^{\lol\lom}(1,2,3,4)
        + \GT^{\lol\lom}_{\lon}(1,2,3,4)\,\qq^{\lon}
        + \GT^{\lol\lom}_{\lon_1\lon_2}(1,2,3,4)\,\qq^{\lon_1}\qq^{\lon_2}}
       {(\qq+\mom_{1,1})^2(\qq+\mom_{1,2})^2(\qq+\mom_{1,3})^2(\qq+\mom_{1,4})^2}
\\
+ \frac{\GT^{\lol\lom}_{\lon_1\lon_2\lon_3}(1,2,3,4)\,\qq^{\lon_1}\qq^{\lon_2}\qq^{\lon_3}}
       {(\qq+\mom_{1,1})^2(\qq+\mom_{1,2})^2(\qq+\mom_{1,3})^2(\qq+\mom_{1,4})^2}
~.
\end{multline}
%
With this observation, we can assign a meaning to the relations of \Equation{Eq391} as follows.
Given the tensor integrals
%
\begin{equation}
\TT^{\lon_{1}\lon_{2}\cdots\lon_{\rr}}(\dd)
=
\int\frac{d^\Dim\qq}{\imag\pi^{\Dim/2}}
\frac{\qq^{\lon_{1}}_{4}\qq^{\lon_{2}}_{4}\cdots\qq^{\lon_{\rr}}_{4}}
     {\prod_{j\in\dd}[(\qq+\mom_{1,j})^2+\imag0]}
~,
\label{Eq512}
\end{equation}
%
for which the numerator only contains $4$-dimensional components of $\qq$, we define the object
%
\begin{equation}
\Gb^{\lol\lom}_{\ixm,\ixn}
=
\sum_{\dd\subset\{\ixm-1,\ixm,\ldots,\ixn\}}
\sum_{\rr=0}^{|\dd|-1}
\GT^{\lol\lom}_{\lon_1\lon_2\cdots\lon_{\rr}}(\dd)
\TT^{\lon_{1}\lon_{2}\cdots\lon_{\rr}}(\dd)
\label{Eq532}
\end{equation}
%
which does not depend on $\qq$, and assign
%
\begin{eqnarray}
\graph{graph05}{41}{20}
=
\frac{-\imag}{\mom_{\ixm,\ixn}^2}
\sum_{\ixk=\ixm}^{\ixn-1}
\VW^{\lom}_{\lon\lor\los}
\,\Gb^{\lon\lor}_{\ixm,\ixk}\,\Aa^{\los}_{\ixk+1,\ixn}
\quad,\quad
\graph{graph06}{41}{20}
=
\frac{-\imag}{\mom_{\ixm,\ixn}^2}
\sum_{\ixk=\ixm}^{\ixn-1}
\VW^{\lom}_{\lon\lor\los}
\,\Aa^{\lon}_{\ixm,\ixk}\,\Gb^{\lor\los}_{\ixk+1,\ixn}
~.
\end{eqnarray}
%
This will lead to a calculation of the one-loop amplitude within the scheme of \cite{Weinzierl:1999xb}.
We cannot use the tensors $\GT^{\lol\lom}_{\lon_1\lon_2\cdots\lon_{\rr}}(\dd)$ directly to define the first line of \Equation{Eq391}, and we will discuss this below.

First, however, we need to answer the question how to calculate the tensors $\GT^{\lol\lom}_{\lon_1\lon_2\cdots\lon_{\rr}}(\dd)$.
Obviously, from \Equation{Eq251} we can derive recursive equations for them.
Writing $\dd=\{\dd',\ixk,\ixn\}$, so the largest two elements of $\dd$ are $\{\ixk,\ixn\}$, we find
%
\begin{eqnarray}
\GT^{\lol\lom}_{\lon_1\lon_2\cdots\lon_{\rr}}(\dd)
&=&
\GT^{\lol\lom}_{\lon_1\lon_2\cdots\lon_{\rr}}(\dd',\ixk,\ixn)
\nonumber\\
&=&
-\imag\,\GT^{\lol\lon}_{\lon_1\lon_2\cdots\lon_{\rr}}(\dd',\ixk)
\left[
 \VV^{\lom}_{\lon\lor}(\mom_{1,\ixk},\mom_{\ixk+1,\ixn})
 \Aa^{\lor}_{\ixk+1,\ixn}
+
\sum_{\ixl=\ixk+1}^{\ixn-1}
\VW^{\lom}_{\lon\lor\los}
\Aa^{\lor}_{\ixk+1,\ixl}\Aa^{\los}_{\ixl+1,\ixn}
\right]
\nonumber\\
&-&
\imag\,\GT^{\lol\lon}_{\lon_1\lon_2\cdots\lon_{\rr-1}}(\dd',\ixk)
\,\VX^{\lom}_{\lon_{\rr}\lon\lor}
  \Aa^{\lor}_{\ixk+1,\ixn}
~,
\label{Eq370}
\end{eqnarray}
%
where the third line is absent for the case $\rr=0$, and the second line is absent for the case $\rr=|\dd|-1$.
As the starting points of the relations we define
%
\begin{equation}
\GT^{\lol\lom}_{}(\ixm)
=
\metric^{\lol\lom}
\quad,\quad
\GT^{\lol\lom}_{\lon}(\ixm)
=
0
~,
\end{equation}
%
for any $\ixm=0,\ldots,\mult$.

Let us address the discussion about the computational complexity in \Section{Sec384} and compare \Equation{Eq370} with \Equation{Eq371}.
The first difference is that \Equation{Eq370} has tensors of rank $\rr$ also on the r.h.s..
Secondly, \Equation{Eq370} involves the contraction with index $\lon$ instead of a simple multiplication.
These differences do not influence the asymptotic computational complexity.
Finally, all objects calculated using the recursive relation are needed in \Equation{Eq532}, and no auxiliary tensors show up whose calculation could influence the asymptotic computational complexity.

In order to deal with the first line of \Equation{Eq391}, we introduce the objects
%
\begin{equation}
\HH^{\lom}_{\ixm,\ixn}(\qq)
=
\VV^{\lom}_{\lon\lor}(-\qq-\mom_{1,\ixm-1},\qq+\mom_{1,\ixn})
\GG^{\lon\lor}_{\ixm,\ixn}(\qq)
~.
\end{equation}
%
With the help of the symbol 
%
\begin{equation}
\VY^{\lom}_{\los\lon\lor}
=
\frac{\imag\gs}{\sqrt{2}}[\;
-2\metric^{\lom}_{\los}\metric_{\lon\lor}
+\metric^{\lom}_{\lor}\metric_{\lon\los}
+\metric^{\lom}_{\lon}\metric_{\lor\los}
\;]
\end{equation}
%
we can separate the $\qq$-dependent part of the $3$-point vertex again and write
%
\begin{equation}
\HH^{\lom}_{\ixm,\ixn}(\qq)
=
\left[
\VV^{\lom}_{\lon\lor}(-\mom_{1,\ixm-1},\mom_{1,\ixn}) + \VY^{\lom}_{\los\lon\lor}\qq^{\los}
\right]
\GG^{\lon\lor}_{\ixm,\ixn}(\qq)
~,
\end{equation}
%
and express
%
\begin{equation}
\HH^{\lom}_{\ixm,\ixn}(\qq)
=
\sum_{\dd\subset\{\ixm-1,\ixm,\ldots,\ixn\}}
\sum_{\rr=0}^{|\dd|}
\HT^{\lom}_{\lon_1\lon_2\cdots\lon_{\rr}}(\dd)
\,\frac{\qq^{\lon_{1}}\qq^{\lon_{2}}\cdots\qq^{\lon_{\rr}}}
     {\prod_{j\in{}\dd}(\qq+\mom_{1,j})^2}
\label{Eq428}
\end{equation}
%
with
%
\begin{equation}
\HT^{\lom}_{\lon_1\lon_2\cdots\lon_{\rr}}(\dd)
=
\VV^{\lom}_{\lon\lor}(-\mom_{1,\ixm-1},\mom_{1,\ixn})
\GT^{\lon\lor}_{\lon_1\lon_2\cdots\lon_{\rr}}(\dd)
+
\VY^{\lom}_{\lon_{\rr}\lon\lor}
\GT^{\lon\lor}_{\lon_1\lon_2\cdots\lon_{\rr-1}}(\dd)
~.
\label{Eq451}
\end{equation}
%
The second term on the r.h.s.\ is absent for the case $\rr=0$ and the first one is absent for the case $\rr=|\dd|$.
Now we simply assign
%
\begin{equation}
\imag\mom_{\ixm,\ixn}^2
\graph{graph04}{41}{6}
=
\Hb^{\lom}_{\ixm,\ixn}
=
\sum_{\dd\subset\{\ixm-1,\ixm,\ldots,\ixn\}}
\sum_{\rr=0}^{|\dd|}
\HT^{\lom}_{\lon_1\lon_2\cdots\lon_{\rr}}(\dd)
\,\TT^{\lon_{1}\lon_{2}\cdots\lon_{\rr}}(\dd)
~.
\label{Eq517}
\end{equation}
%

We can now write down the diagrammatic relation \Equation{Eq240} explicitly.
Denoting a one-loop off-shell current by $\BB^{\lom}_{\ixm,\ixn}$, we have
%
\begin{multline}
\BB^{\lom}_{\ixm,\ixn}
=
\frac{-\imag}{\mom_{\ixm,\ixn}^2}
\Bigg\{
\sum_{\ixk=\ixm}^{\ixn-1}
\VV^{\lom}_{\lon\lor}(\mom_{\ixm,\ixk},\mom_{\ixk+1,\ixn})
\Big[
 \BB^{\lon}_{\ixm,\ixk}\,\Aa^{\lor}_{\ixk+1,\ixn}
+\Aa^{\lon}_{\ixm,\ixk}\,\BB^{\lor}_{\ixk+1,\ixn}
\Big]
\\
+
\sum_{\ixk=\ixm}^{\ixn-2}
\sum_{\ixl=\ixk+1}^{\ixn-1}
\VW^{\lom}_{\lon\lor\los}
\Big[
  \BB^{\lon}_{\ixm,\ixk}\,\Aa^{\lor}_{\ixk+1,\ixl}\,\Aa^{\los}_{\ixl+1,\ixn}
+ \Aa^{\lon}_{\ixm,\ixk}\,\BB^{\lor}_{\ixk+1,\ixl}\,\Aa^{\los}_{\ixl+1,\ixn}
+ \Aa^{\lon}_{\ixm,\ixk}\,\Aa^{\lor}_{\ixk+1,\ixl}\,\BB^{\los}_{\ixl+1,\ixn}
\Big]
\\
+ \Hb^{\lom}_{\ixm,\ixn}
+ \sum_{\ixk=\ixm}^{\ixn-1}
  \VW^{\lom}_{\lon\lor\los}
  \Big[
    \Gb^{\lon\lor}_{\ixm,\ixk}\,\Aa^{\los}_{\ixk+1,\ixn}
  + \Aa^{\lon}_{\ixm,\ixk}\,\Gb^{\lor\los}_{\ixk+1,\ixn}
  \Big]
\Bigg\}
~.
\label{Eq745}
\end{multline}
%
So the program to calculate these one-loop off-shell currents is to
\begin{enumerate}
\item calculate the tree-level off-shell currents $\Aa^{\lom}_{\ixm,\ixn}$ with recursive equation \Equation{Eq137};
\item calculate the tensor integrals $\TT^{\lon_1\lon_2\cdots\lon_{\rr}}(\dd)$ defined in \Equation{Eq512};
\item calculate the tensors $\GT^{\lor\lom}_{\lon_1\lon_2\cdots\lon_{\rr}}(\dd)$ using \Equation{Eq370} and the tensors $\HT^{\lom}_{\lon_1\lon_2\cdots\lon_{\rr}}(\dd)$ using \Equation{Eq451};
\item calculate the objects $\Hb^{\lom}_{\ixm,\ixn}$ following \Equation{Eq517} and $\Gb^{\lor\lom}_{\ixm,\ixn}$ following \Equation{Eq532};
\item solve \Equation{Eq745} recursively.
\end{enumerate}
The one-loop amplitude finally is given by
%
\begin{equation}
\Amp^{(1)}_{\mult}
=
\metric_{\lon\lom}\,\Pol^{\lon}_{\mult}\,\mom_{\mult}^2\BB^{\lom}_{1,\mult-1}
~.
\end{equation}
%

%% file: ghostandr2.tex
In order to arrive at gauge-invariant amplitudes, ghost contributions have to be included.
Also, one could want to include quark loops.
Then, \Equation{Eq240} has to be extended to
%
\begin{multline}
\graph{graph07}{36}{6}
=
  \graph{graph08}{41}{20}
+ \graph{graph09}{41}{20}
+ \graph{graph10}{51}{26}
+ \graph{graph12}{51}{26}
+ \graph{graph11}{51}{26}
\\
+ \graph{graph04}{41}{6}
+ \graph{graph05}{41}{20}
+ \graph{graph06}{41}{20}
+ 2\graph{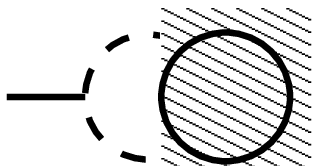}{42}{6.5}
- \frac{1}{\ncol}\graph{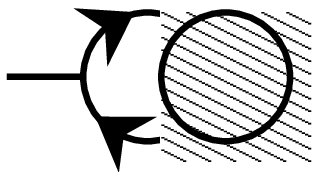}{42}{11}
\quad.
\end{multline}
%
The factor $2$ for the ghost loop is needed to arrive at gauge-invariant ordered amplitudes.
The calculation of the ghost-tensors is rather trivial once the calculation of the gluonic ones are understood.
We prefer to focus the attention on the quark loops.
The main difference is that for these, the $\qq$-dependence of the numerators in the off-shell currents with an auxiliary quark comes from the quark propagators instead of the vertices.
The off-shell currents, or ``off-shell spinors'', with an auxiliary quark satisfy
%
\begin{equation}
\UU^{\spi,\spj}_{\ixm,\ixn}(\qq)
=
\graph{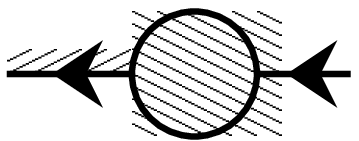}{48}{6}
=
\graph{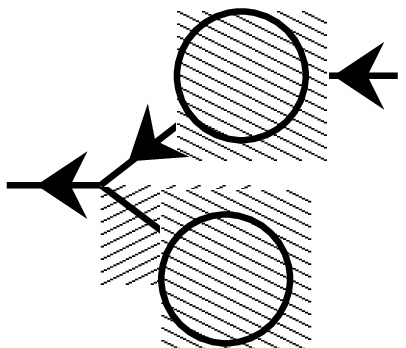}{56}{20}
\quad,
\end{equation}
%
where $\spi,\spj$ denote the spinor-indices.
More explicitly, the relation can be written as
%
\begin{equation}
\UU^{\spi,\spj}_{\ixm,\ixn}(\qq)
=
\frac{\imag}{(\qq+\mom_{1,\ixn})^2}
\sum_{\ixk=\ixm-1}^{\ixn-1}
\,(\Gamma_{\lom\lon})^{\spi}_{\spk}
\,(\qq+\mom_{1,\ixn})^{\lom}
\,\UU_{\ixm,\ixk}^{\spk,\spj}(\qq)
\,\Aa_{\ixk+1,\ixn}^{\lon}
\end{equation}
%
where implicit summation also over spinor-indices is understood.
Here, we introduced the matrix
%
\begin{equation}
\Gamma_{\lom\lon}
=
\frac{\imag\gs}{\sqrt{2}}\,\gamma_{\lom}\gamma_{\lon}
~.
\end{equation}
%
The $\qq$-dependent part on the r.h.s.\ can be isolated even more straightforwardly than in the gluon case.
We write
%
\begin{equation}
\UU^{\spi,\spj}_{\ixm,\ixn}(\qq)
=
\sum_{\dd\subset\{\ixm-1,\ixm,\ldots,\ixn\}}
\sum_{\rr=0}^{|\dd|-1}
\UT^{\spi,\spj}_{\lon_1\lon_2\cdots\lon_{\rr}}(\dd)
\,\frac{\qq^{\lon_{1}}\qq^{\lon_{2}}\cdots\qq^{\lon_{\rr}}}
       {\prod_{j\in{}\dd}(\qq+\mom_{1,j})^2}
\end{equation}
%
and find that
%
\begin{eqnarray}
\UT^{\spi,\spj}_{\lon_1\lon_2\cdots\lon_{\rr}}(\dd)
=
\UT^{\spi,\spj}_{\lon_1\lon_2\cdots\lon_{\rr}}(\dd',\ixk,\ixn)
&=&
\imag\,\UT^{\spk,\spj}_{\lon_1\lon_2\cdots\lon_{\rr}}(\dd',\ixk)
\,(\Gamma_{\lom\lor})^{\spi}_{\spk}
\,\mom_{1,\ixn}^{\lom}
\,\Aa^{\lor}_{\ixk+1,\ixn}
\nonumber\\
&+&
\imag\,\UT^{\spk,\spj}_{\lon_1\lon_2\cdots\lon_{\rr-1}}(\dd',\ixk)
\,(\Gamma_{\lon_{\rr}\lor})^{\spi}_{\spk}
\,\Aa^{\lor}_{\ixk+1,\ixn}
~.
\end{eqnarray}
%
The contribution from the quark loops to the integrands of the one-loop gluon off-shell currents, without the the factor $-1/\ncol$, are given by
%
\begin{equation}
\FF^{\lom}_{\ixm,\ixn}(\qq)
=
(\qq+\mom_{1,\ixm-1})^{\lon}
\,({\Gamma_{\lon}}^{\lom})_{\spj,\spi}
\,\UU^{\spi,\spj}_{\ixm,\ixn}(\qq)
~.
\end{equation}
%
Remember that $\UU^{\spi,\spj}_{\ixm,\ixn}(\qq)$ already contains the denominator of the propagator factor on the r.h.s..
Now we write
%
\begin{equation}
\FF^{\lom}_{\ixm,\ixn}(\qq)
=
\sum_{\dd\subset\{\ixm-1,\ixm,\ldots,\ixn\}}
\sum_{\rr=0}^{|\dd|}
\FT^{\lom}_{\lon_1\lon_2\cdots\lon_{\rr}}(\dd)
\,\frac{\qq^{\lon_{1}}\qq^{\lon_{2}}\cdots\qq^{\lon_{\rr}}}
     {\prod_{j\in{}\dd}(\qq+\mom_{1,j})^2}
\end{equation}
%
with
%
\begin{eqnarray}
\FT^{\lom}_{\lon_1\lon_2\cdots\lon_{\rr}}(\dd)
&=&
\mom_{1,\ixm-1}^{\lon}
\,({\Gamma_{\lon}}^{\lom})_{\spj,\spi}
\,\UT^{\spi,\spj}_{\lon_1\lon_2\cdots\lon_{\rr}}(\dd',\ixk)
\nonumber\\
&+&
({\Gamma_{\lon_{\rr}}}^{\lom})_{\spj,\spi}
\,\UT^{\spi,\spj}_{\lon_1\lon_2\cdots\lon_{\rr-1}}(\dd',\ixk)
~.
\end{eqnarray}
%
The integrated contributions from the quark loops to the one-loop gluon off-shell currents are then given by
%
\begin{equation}
-\imag\mom_{\ixm,\ixn}^2
\graph{graph53}{42}{11}
=
\sum_{\dd\subset\{\ixm-1,\ixm,\ldots,\ixn\}}
\sum_{\rr=0}^{|\dd|}
\FT^{\lom}_{\lon_1\lon_2\cdots\lon_{\rr}}(\dd)
\,\TT^{\lon_{1}\lon_{2}\cdots\lon_{\rr}}(\dd)
~.
\end{equation}
%
%
\subsection{The $\Rtwo$-term}
%
Finally, to obtain gauge-invariant results, the $\Rtwo$-term~\cite{Ossola:2008xq,Draggiotis:2009yb} has to be included.
The necessary extra vertices are given by
%
\begin{eqnarray}
\PPR^{\lom}_{\lon}(\mom)
&=&
\frac{\gs^2}{3}(1/2+\lambda_{\mathrm{HV}}+\nfer/\ncol)
\mom^2\metric^{\lom}_{\lon}
\\
\VVR^{\lom}_{\lon\lor}(\mom_{1},\mom_{2})
&=&
\frac{-\gs^3}{3\sqrt{2}}(7/4 + \lambda_{\mathrm{HV}} + 2\nfer/\ncol)
[\; (\mom_{1}-\mom_{2})^{\lom}\metric_{\lon\lor}
   +2\mom_{2\lon}\,\metric_{\lor}^{\lom}
   -2\mom_{1\lor}\,\metric_{\lon}^{\lom}\;]
\\
\VWR^{\lom}_{\lon\lor\los}
&=&
\frac{-\gs^4}{6}
[\;
   (6+2\lambda_{\mathrm{HV}}+5\nfer/\ncol)\metric^{\lom}_{\lor}\,\metric_{\lon\los}
\nonumber\\
 &&\qquad-(5/2+\lambda_{\mathrm{HV}}+3\nfer/\ncol)
    (\metric^{\lom}_{\lon}\,\metric_{\lor\los}+\metric^{\lom}_{\los}\,\metric_{\lor\lon})
\;]
~,
\end{eqnarray}
%
where $\nfer$ is then number of quarks, and $\lambda_{\mathrm{HV}}=1$ in the 't~Hooft-Veltman scheme and $\lambda_{\mathrm{HV}}=0$ in the FDH scheme.
The off-shell currents $\RR^{\lom}_{\ixm,\ixn}$ containing graphs with exactly $1$ such vertex satisfy the recursive equation
%
\begin{multline}
\RR^{\lom}_{\ixm,\ixn}
=
\frac{-\imag}{\mom_{\ixm,\ixn}^2}
\Bigg\{
\sum_{\ixk=\ixm}^{\ixn-1}
\VV^{\lom}_{\lon\lor}(\mom_{\ixm,\ixk},\mom_{\ixk+1,\ixn})
\Big[
 \RR^{\lon}_{\ixm,\ixk}\,\Aa^{\lor}_{\ixk+1,\ixn}
+\Aa^{\lon}_{\ixm,\ixk}\,\RR^{\lor}_{\ixk+1,\ixn}
\Big]
\\
+
\sum_{\ixk=\ixm}^{\ixn-2}
\sum_{\ixl=\ixk+1}^{\ixn-1}
\VW^{\lom}_{\lon\lor\los}
\Big[
  \RR^{\lon}_{\ixm,\ixk}\,\Aa^{\lor}_{\ixk+1,\ixl}\,\Aa^{\los}_{\ixl+1,\ixn}
+ \Aa^{\lon}_{\ixm,\ixk}\,\RR^{\lor}_{\ixk+1,\ixl}\,\Aa^{\los}_{\ixl+1,\ixn}
+ \Aa^{\lon}_{\ixm,\ixk}\,\Aa^{\lor}_{\ixk+1,\ixl}\,\RR^{\los}_{\ixl+1,\ixn}
\Big]
\\
+
\PPR^{\lom}_{\lon}(\mom_{\ixm,\ixn}^2)\Aa^{\lon}_{\ixm,\ixn}
+
\sum_{\ixk=\ixm}^{\ixn-1}
\VVR^{\lom}_{\lon\lor}(\mom_{\ixm,\ixk},\mom_{\ixk+1,\ixn})
\Aa^{\lon}_{\ixm,\ixk}\,\Aa^{\lor}_{\ixk+1,\ixn}
\\
+
\sum_{\ixk=\ixm}^{\ixn-2}
\sum_{\ixl=\ixk+1}^{\ixn-1}
\VWR^{\lom}_{\lon\lor\los}
\Aa^{\lon}_{\ixm,\ixk}\,\Aa^{\lor}_{\ixk+1,\ixl}\,\Aa^{\los}_{\ixl+1,\ixn}
\Bigg\}
~.
\end{multline}
%
The $\Rtwo$-term is then given by
%
\begin{equation}
\Rtwo
=
\metric_{\lon\lom}\,\Pol^{\lon}_{\mult}\,\mom_{\mult}^2\RR^{\lom}_{1,\mult-1}
~.
\end{equation}

%% file: results.tex
The presented algorithm has been implemented in a {\tt Fortran77}\ program which, first of all, reproduces all the numeric results given in \cite{Giele:2008bc} for multi-gluon amplitudes up to $10$ gluons to at least $4$ decimals precision (\Appendix{App1}).
Results involving one massless quark-loop are presented in \Appendix{App2}.

Secondly an analysis of the accuracy like in \cite{Giele:2008bc} and \cite{Lazopoulos:2008ex} has been performed, which makes use of the existence of a simple formula for the divergent part of color-ordered one-loop gluon amplitudes within dimensional regularization~\cite{Giele:1991vf}.
It is given in~\cite{Giele:2008bc} as
%
\begin{equation}
\Amp^{(\mathrm{poles})}_{\mult}
=
\frac{(4\pi)^\epsilon}{16\pi^2}
\frac{\Gamma(1+\epsilon)\Gamma^2(1-\epsilon)}{\Gamma(1-2\epsilon)}
\left\{
\frac{-\mult}{\epsilon^2}
+
\frac{1}{\epsilon}\left[
\sum_{\ixm=1}^{\mult}\ln\left(\frac{-\mom_{\ixm,\ixm+1}^2}{\mu^2}\right)
-\frac{11}{3}
\right]
\right\}
\Amp^{(\mathrm{tree})}_{\mult}
\label{Result23}
~,
\end{equation}
%
where $\epsilon=(4-\Dim)/2$ and $\Dim$ is the dimension, $\mu$ is the dimensional scale.
Since the divergent part of the one-loop amplitude should also be obtained by using the divergent parts of the initial scalar functions as the starting points in the calculation of the tensor integrals, there is the opportunity to compare the two and assess the accuracy of the latter.
The left of \Figure{Fig1} gives the distribution of the quantity
%
\begin{equation}
^{10}\log\left|\frac{\Amp^{(1/\epsilon)}_{\mult}(\ref{Result23})
                    -\Amp^{(1/\epsilon)}_{\mult}}
                    {\Amp^{(1/\epsilon)}_{\mult}(\ref{Result23})}\right|
~,
\label{Result47}
\end{equation}
%
where $\Amp^{(1/\epsilon)}_{\mult}(\ref{Result23})$ refers to the coefficient of $1/\epsilon$ in \Equation{Result23}, and  $\Amp^{(1/\epsilon)}_{\mult}$ refers to this coefficient calculated with the program presented in this write-up.
The distribution is obtained from a large sample of uniformly distributed phase-space points with the same kinematical cuts as in \cite{Giele:2008bc,Lazopoulos:2008ex} being
%
\begin{equation}
|\eta_{\ixm}|<3
\quad,\quad
\mom_{\perp,\ixm} > 0.01\sqrt{s}
\quad,\quad
\sqrt{|\phi_{\ixm}-\phi_{\ixn}|^2+|\eta_{\ixm}-\eta_{\ixn}|^2} > 0.4
~,
\end{equation}
%
where $\eta_{\ixm}$ is the rapidity of gluon $\ixm$, $\mom_{\perp,\ixm}$ its transverse momentum and $\eta_{\ixm}$ its azimuthal angle, all with respect to the axis of the incoming gluons.
Also helicity configurations where sampled uniformly distributed, only avoiding configurations for which the tree-level amplitude vanishes.
The distributions for a total number of $6$, $8$ and $10$ gluons are shown for calculations at the double precision level.
We see a behavior compatible with \cite{Giele:2008bc} at the double precision level, and slightly better than \cite{Lazopoulos:2008ex} at double precision level.
The lower graph at the left of \Figure{Fig1} shows the same distributions in a log scale for the y-axis in order to highlight the right tail.

Obviously, for a small part of the phase space points the accuracy becomes unacceptably bad.
Results of re-evaluation for these at higher precision are given on the left of \Figure{Fig2}.
Presented are the right tail of the distribution for $\mult=10$ starting from $-4$, and distributions obtained with the same set of phase space points for different options of applying quadruple precision arithmetic.
Evaluating the tensor integrals only (but including the scalar functions) at quadruple precision, the tail is shifted to the left to a large extend, but still phase space points may show up leading to an unacceptably bad accuracy.
The more expensive option of evaluating everything at the quadruple precision level, however, moves the whole tail below $-16$.

Another hint at the accuracy of the program can be given by the extend at which gauge invariance is satisfied.
A one-loop amplitude should vanish whenever the polarization vector of any of the external gluons is replaced by the momentum of that gluon.
The quantity
%
\begin{equation}
^{10}\log\left(
\left|\frac{\Re\,\Amp^{(1)}_{\mult}(\varepsilon_{\ixm}\leftarrow\mom_{\ixm})}
           {\Re\,\Amp^{(1)}_{\mult}}\right|
+
\left|\frac{\Im\,\Amp^{(1)}_{\mult}(\varepsilon_{\ixm}\leftarrow\mom_{\ixm})}
           {\Im\,\Amp^{(1)}_{\mult}}\right|
\right)
\label{Result83}
\end{equation}
%
may serve as a measure of the number of decimals being eliminated by performing such a replacement in a numerical calculation, which then again may give an estimate of the accuracy.
The right of \Figure{Fig1} presents the distribution of (the finite part of) this quantity, obtained from the same sample of phase-space points as before, now however including helicity configurations for which the tree-level amplitude vanishes.
The distributions are compatible with the ones on the left.
The lower graph at the rights show the same distributions again in a log scale for the y-axis.
The right tails behave similarly to the ones left, and the same holds for the distributions in \Figure{Fig2}.

\myFigure{
\epsfig{figure=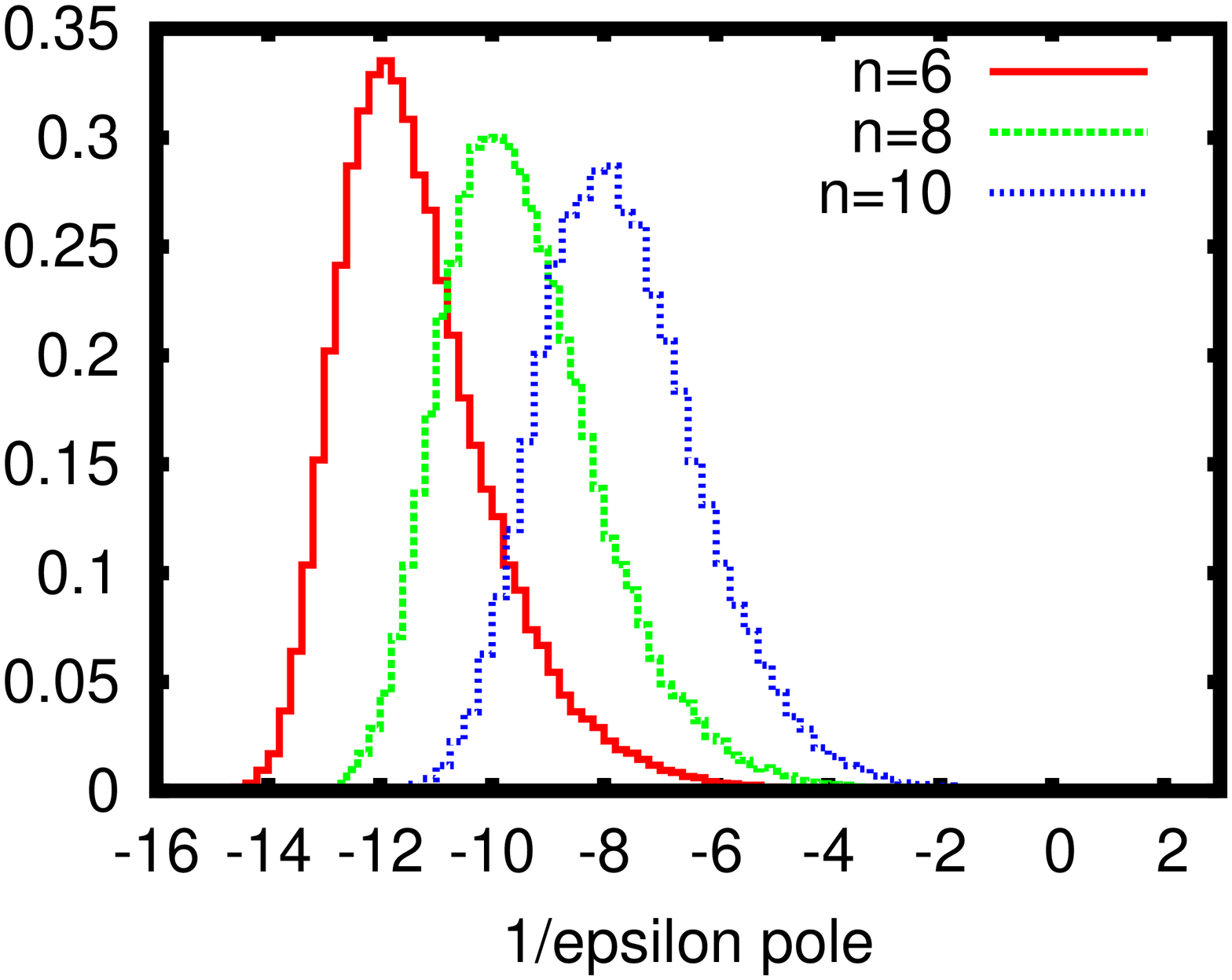,width=0.33\linewidth,angle=270}
\epsfig{figure=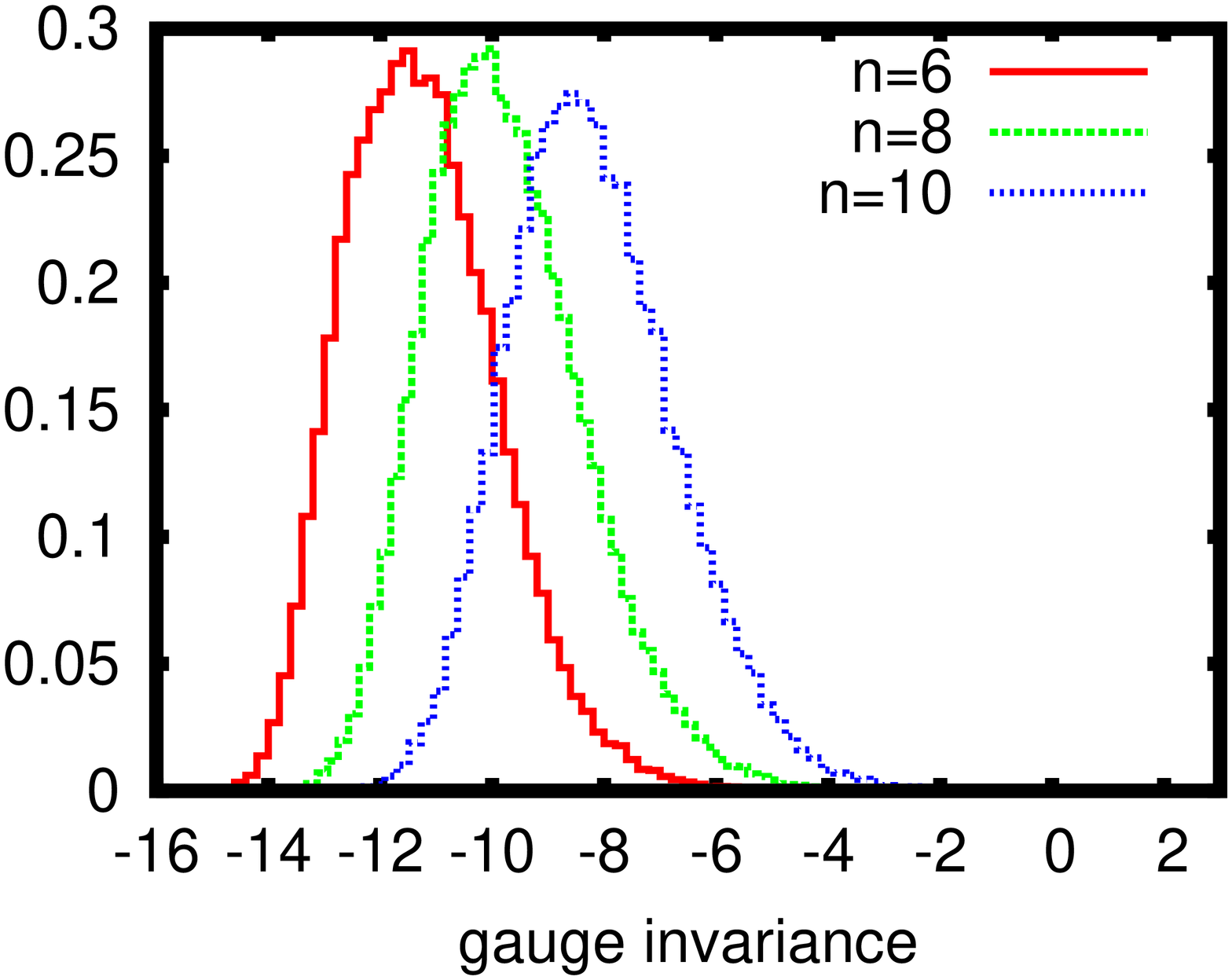,width=0.33\linewidth,angle=270}\\
\epsfig{figure=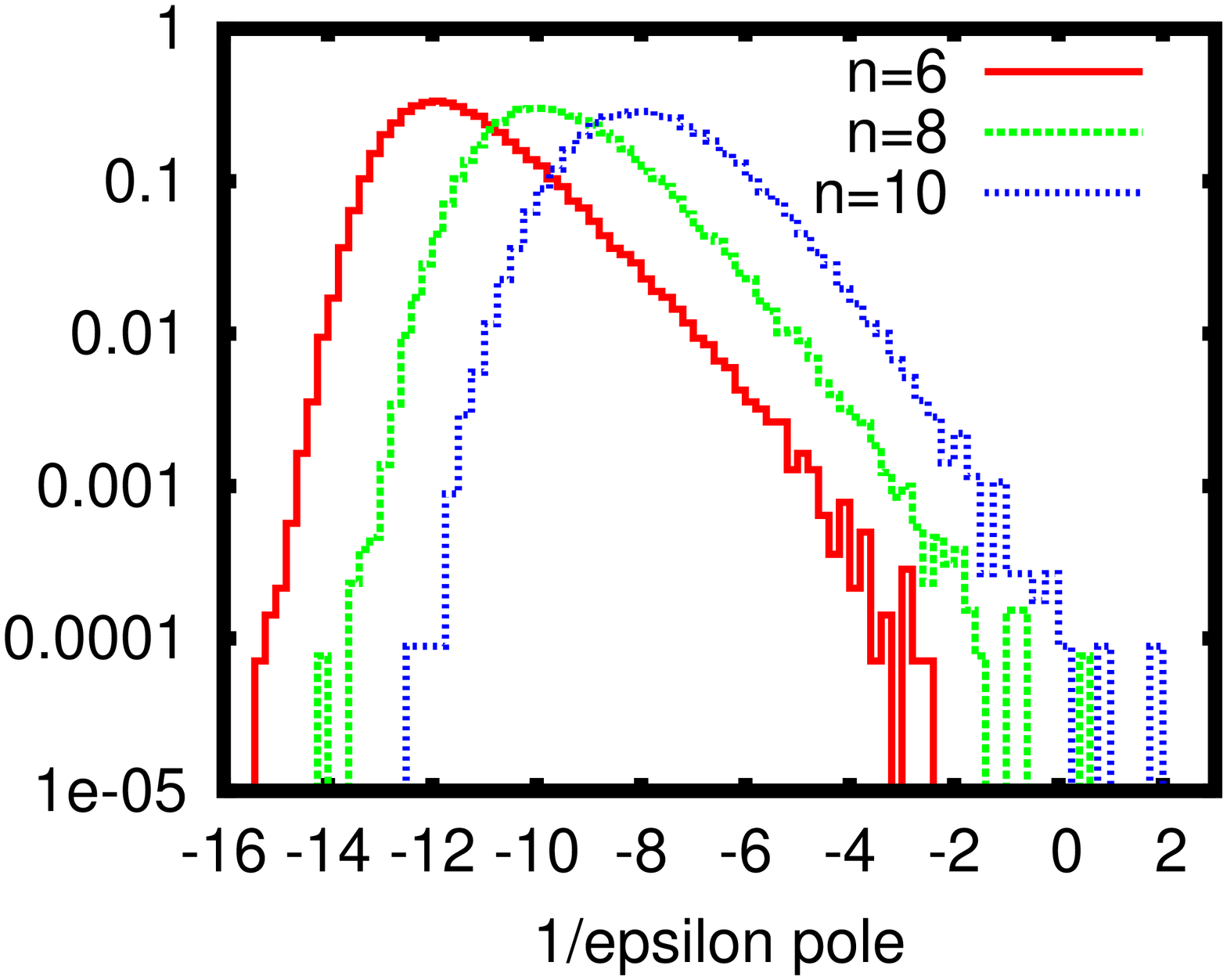,width=0.33\linewidth,angle=270}
\epsfig{figure=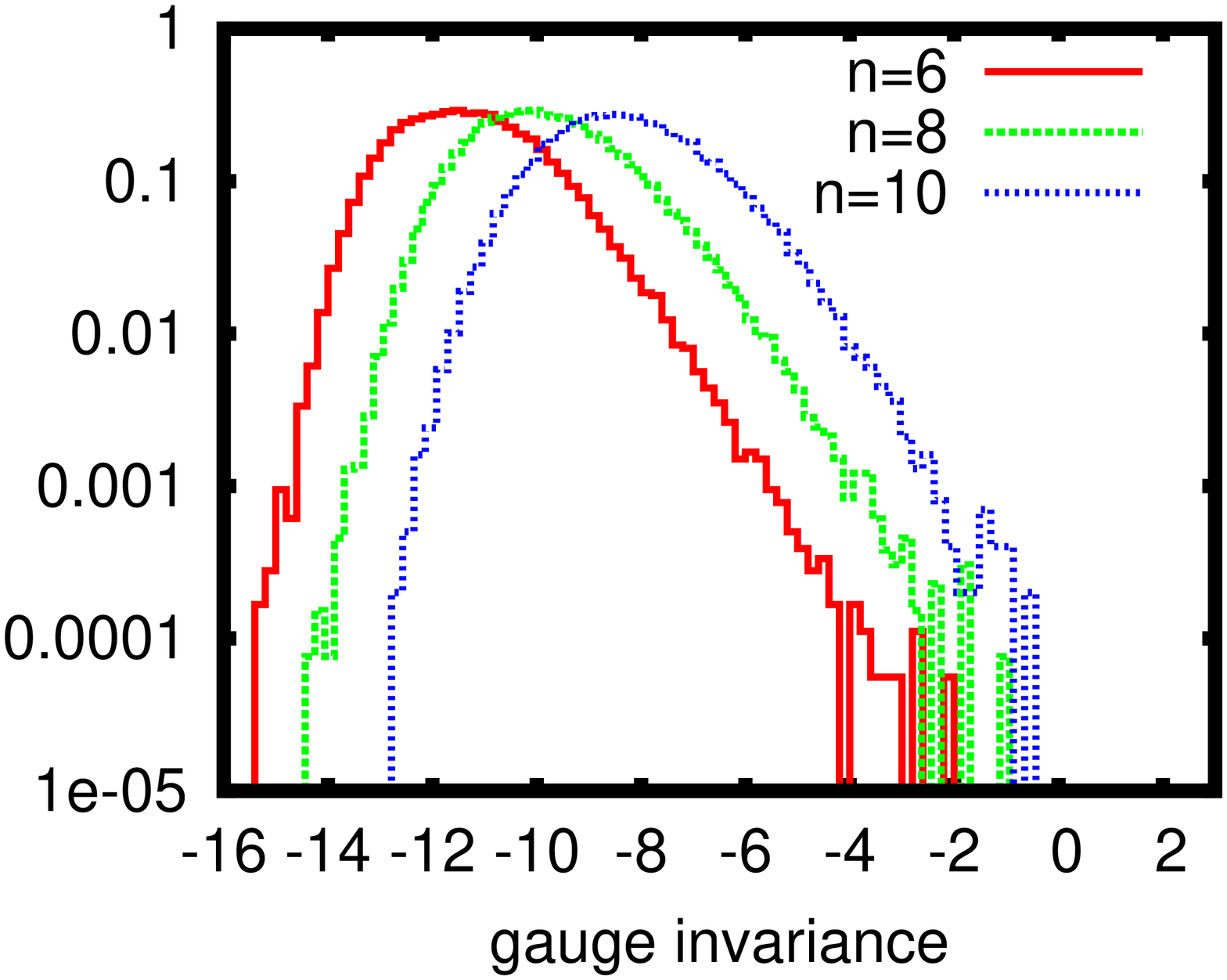,width=0.33\linewidth,angle=270}
\caption{The distribution of the quantity in \Equation{Result47} (left) and \Equation{Result83} (right) for calculations at the double precision level. The lower graphs represent the same distributions as the upper ones, but with a logarithmic scale for the y-axis.}
\label{Fig1}
}
\myFigure{
\epsfig{figure=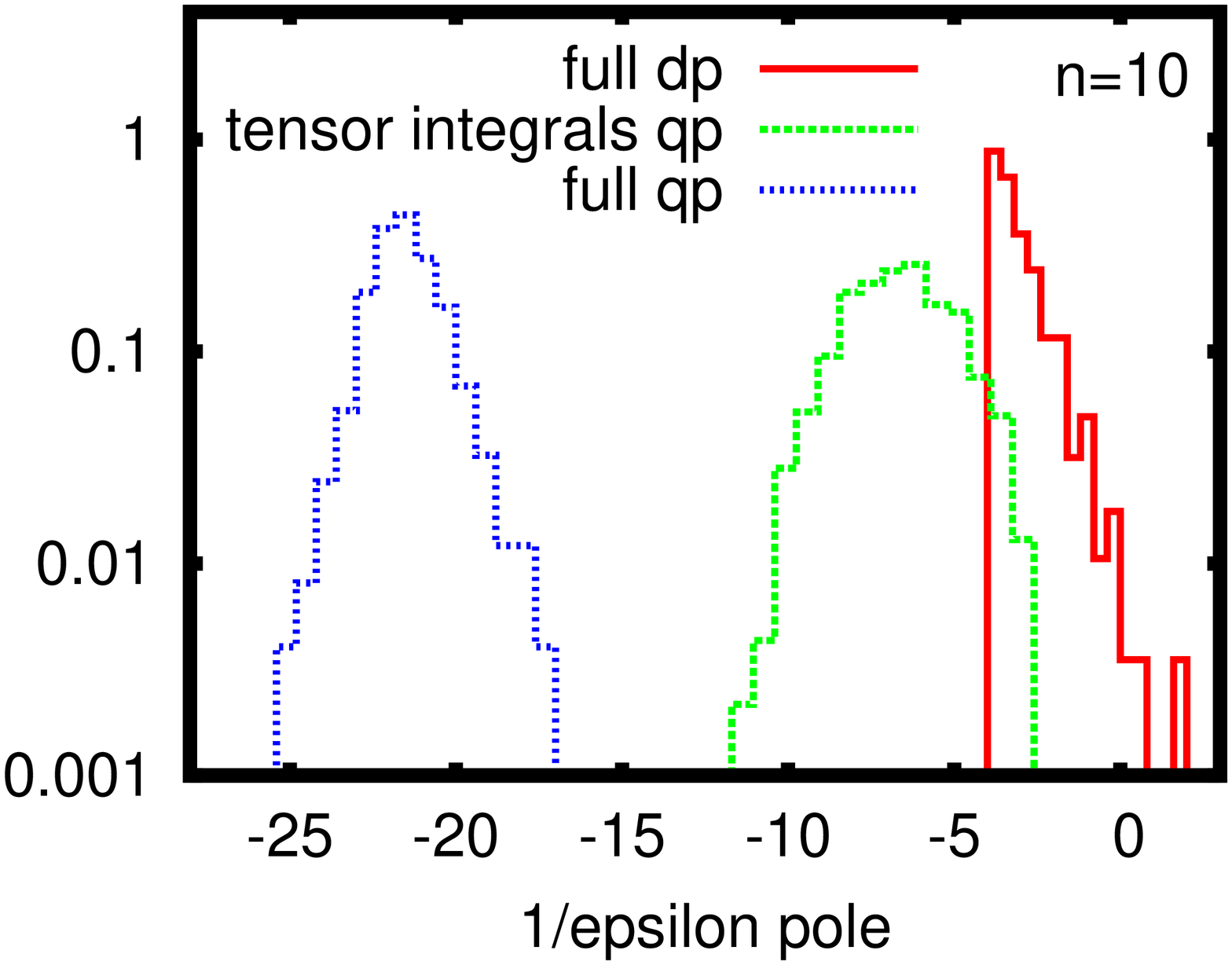,width=0.33\linewidth,angle=270}
\epsfig{figure=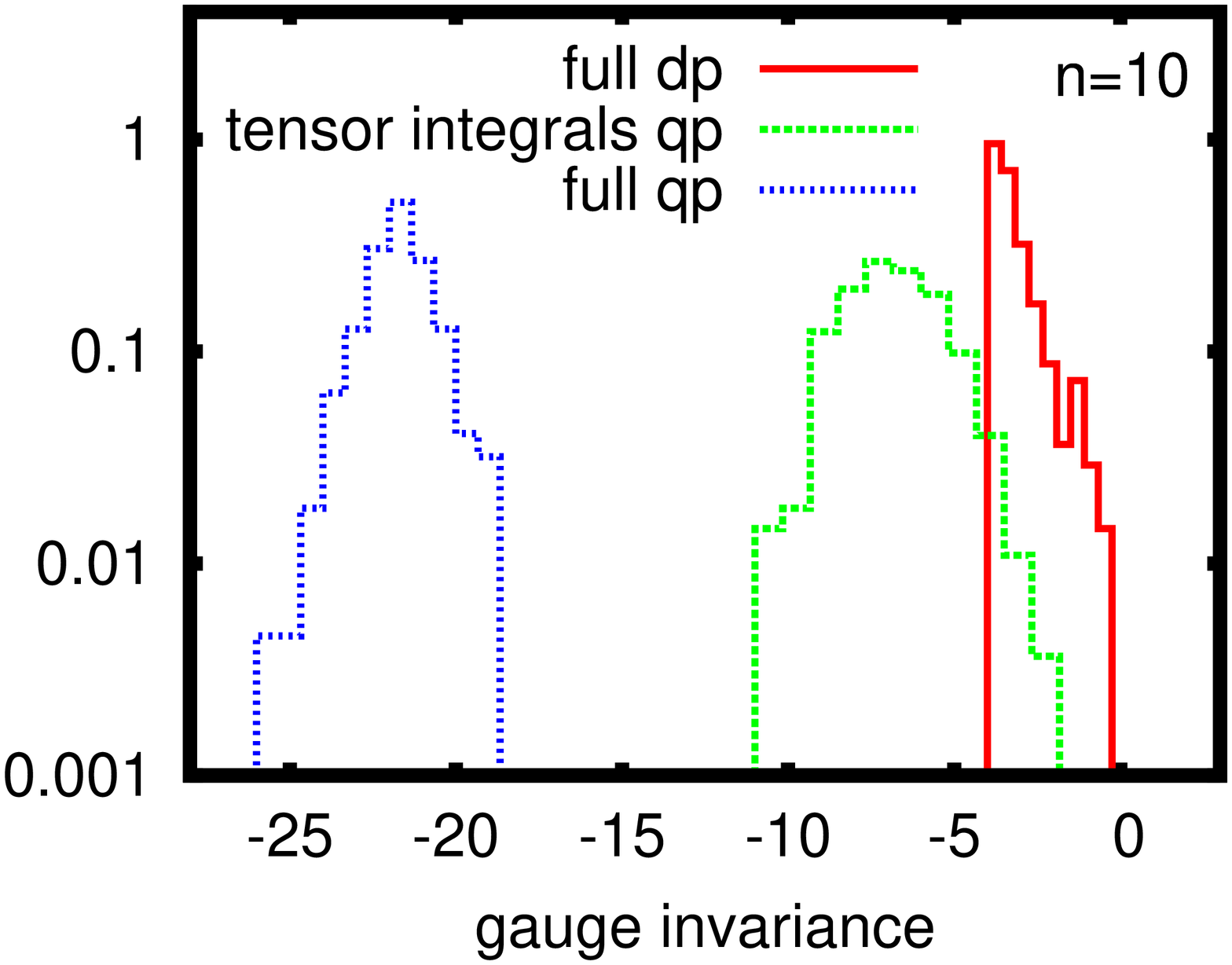,width=0.33\linewidth,angle=270}
\caption{The right tail of the distributions for $n=10$ of \Figure{Fig1} (evaluated at double precision level), and the distributions for the same phase space points when the tensor integrals are evaluated at quadruple precision, and when everything is evaluated at quadruple precision.}
\label{Fig2}
}

In \Table{Tab1} the typical cpu-times $t_{\mathrm{loop}}$ are given needed for $1$ evaluation of the one-loop amplitude for a number of gluons from $4$ to $10$ on a 2.80GHz Intel Xeon processor.
They are determined by taking the average over the evaluations for a large number of different phase-space points.
The numbers are roughly comparable with those in \cite{Giele:2008bc,Lazopoulos:2008ex}, but do show a worse behavior as function of the number of external particles.
The numbers for $t_{\mathrm{loop}}/(\mult^42^{\mult})$ seem to converge, supporting the statements in \Section{Sec384} about the computational complexity being $\Ord(\mult^42^{\mult})$.
The numbers for $t_{\mathrm{loop}}/t_{\mathrm{tree}}$, where $t_{\mathrm{tree}}$ is the cpu-time needed for $1$ evaluation of the tree-level amplitude, give a machine-independent measure of the computational cost.
Notice that they increase as $2^\mult$, consistent with the computational complexities of $\Ord(\mult^42^\mult)$ for the one-loop amplitude, and $\Ord(\mult^4)$ for the tree-level amplitude.
\myTable{
\begin{tabular}{|c||c|c|c|c|c|c|c|}
\hline
$\mult$
& 4 &   5 &   6 &    7 &    8 &    9 & 10
\\\hline\hline
$t_{\mathrm{loop}}(\mathrm{ms})$
& 2.762  & 10.15  & 34.37  & 109.8  & 335.1  & 965.2  & 2744
\\\hline
$ t_{\mathrm{loop}}(\mu\mathrm{s})/(\mult^42^{\mult}) $
& 0.6744 & 0.5077 & 0.4144 & 0.3573 & 0.3196 & 0.2873 & 0.2680
\\\hline
$t_{\mathrm{loop}}/t_{\mathrm{tree}}/10^3$
& 0.2990 & 0.6102 & 1.180  & 2.244  & 4.104  & 7.919  & 15.17
\\\hline
\end{tabular}
\caption{Typical cpu-times needed for $1$ evaluation of the one-loop amplitude on a 2.80GHz Intel Xeon processor.}
\label{Tab1}
}
%
%
%
%
%
%
%
%
%

%% file: conclusion.tex
An algorithm was presented to calculate multi-gluon one-loop amplitudes using tensor integrals, which was shown to be competitive with existing programs using the unitarity-approach up to a number of $10$ gluons.
It  uses universal recursive relations for tensor integrals, independent of the amplitude being calculated.
It also uses recursive relations for ordered gluon amplitudes, however in this respect it can straightforwardly be generalized to any field theory by extending known recursive relations at tree-level \cite{Caravaglios:1995cd,Draggiotis:1998gr,Kanaki:2000ey} to the one-loop level, as was done for the ordered gluon amplitudes in this write-up.
%

%% file: appendixA.tex
This appendix contains the reproduction of the explicit numeric results in \cite{Giele:2008bc} up to a number of $10$ gluons.
The phase-space points at which the amplitudes were evaluated can be found in \cite{Giele:2008bc}.
Presented is the absolute value of the coefficients of $\epsilon^{-2}$, $\epsilon^{-1}$ and $\epsilon^{0}$ of the amplitudes.
Results labelled with ``tree'' give the latter value for the tree-level amplitude.
The upper-left box in each table gives the helicity configuration.
The results from \cite{Giele:2008bc} are labelled with ``\GZ''.

\subsection{$\mult=6$}

\begin{center}{\scriptsize\begin{tabular}{|c|c|c|c|}
\hline
       ${++++++}$    & $\epsilon^{-2}$        & $\epsilon^{-1}$        & $\epsilon^{0}$
\\\hline
   tree            &                        &                        & 0.2215923815877299E-14
\\ tree \GZ        &                        &                        & 0.1767767365814634E-14
\\ unit \GZ        & 0.0000000000000000E+00 & 0.0000000000000000E+00 & 0.5298064836438550E+00
\\ num \GZ         & 0.1060660419488780E-13 & 0.3813284749527035E-13 & 0.5298064836612950E+00
\\ tnsr            & 0.1349521124077591E-10 & 0.7792283713971911E-10 & 0.5298064837199270E+00
\\\hline
\end{tabular}}\end{center}                                                         

\begin{center}{\scriptsize\begin{tabular}{|c|c|c|c|}
\hline
     ${-+++++}$      & $\epsilon^{-2}$        & $\epsilon^{-1}$        & $\epsilon^{0}$
\\\hline
   tree            &                        &                        & 0.1947124136075826E-13
\\ tree \GZ        &                        &                        & 0.3963158957208070E-13
\\ unit \GZ        & 0.1011255761241711E-10 & 0.6753625348984687E-09 & 0.3259967043518990E+01
\\ num \GZ         & 0.2377895374324842E-12 & 0.8549005883762705E-12 & 0.3259967054272360E+01
\\ tnsr            & 0.6401972138850929E-10 & 0.1571282783053386E-08 & 0.3259967053973336E+01
\\\hline
\end{tabular}}\end{center}                                                         

\begin{center}{\scriptsize\begin{tabular}{|c|c|c|c|}
\hline
     ${--++++}$      & $\epsilon^{-2}$        & $\epsilon^{-1}$        & $\epsilon^{0}$
\\\hline
   tree            &                        &                        & 0.2849128165044324E+02
\\ tree \GZ        &                        &                        & 0.2849128165044320E+02
\\ unit \GZ        & 0.1709476899026590E+03 & 0.6145908783763959E+03 & 0.1373747535008540E+04
\\ num \GZ         & 0.1709476899026590E+03 & 0.6145908783763970E+03 & 0.1373747535008280E+04
\\ tnsr            & 0.1709476899026128E+03 & 0.6145908783750691E+03 & 0.1373747535007119E+04
\\\hline
\end{tabular}}\end{center}                                                         

\begin{center}{\scriptsize\begin{tabular}{|c|c|c|c|}
\hline
     ${-+-+-+}$      & $\epsilon^{-2}$        & $\epsilon^{-1}$        & $\epsilon^{0}$
\\\hline
   tree            &                        &                        & 0.3138715395008085E+01
\\ tree \GZ        &                        &                        & 0.3138715395008080E+01
\\ unit \GZ        & 0.1883229237004670E+02 & 0.6770582934748300E+02 & 0.1510439503289600E+03
\\ num \GZ         & 0.1883229237004850E+02 & 0.6770582928695769E+02 & 0.1510439503379470E+03
\\ tnsr            & 0.1883229236988423E+02 & 0.6770582928667301E+02 & 0.1510439503340157E+03
\\\hline
\end{tabular}}\end{center}                                                         

\begin{center}{\scriptsize\begin{tabular}{|c|c|c|c|}
\hline
     ${+-+-+-}$      & $\epsilon^{-2}$        & $\epsilon^{-1}$        & $\epsilon^{0}$
\\\hline
   tree            &                        &                        & 0.3138715395008091E+01
\\ tree \GZ        &                        &                        & 0.3138715395008080E+01
\\ unit \GZ        & 0.1883229237005540E+02 & 0.6770582928570479E+02 & 0.1537801015298360E+03
\\ num \GZ         & 0.1883229237004850E+02 & 0.6770582928695769E+02 & 0.1537801014159860E+03
\\ tnsr            & 0.1883229236987900E+02 & 0.6770582928717265E+02 & 0.1537801016156306E+03
\\\hline
\end{tabular}}\end{center}                                                         

\subsection{$\mult=7$}

\begin{center}{\scriptsize\begin{tabular}{|c|c|c|c|}
\hline
     ${+++++++}$     & $\epsilon^{-2}$        & $\epsilon^{-1}$        & $\epsilon^{0}$
\\\hline
   tree            &                        &                        & 0.2703922191931054E-14
\\ tree \GZ        &                        &                        & 0.0000000000000000E+00
\\ unit \GZ        & 0.0000000000000000E+00 & 0.1256534542409480E-09 & 0.3101695329720260E+00
\\ anly \GZ        & 0.0000000000000000E+00 & 0.1250170111559883E-14 & 0.3101695334831830E+00
\\ tnsr            & 0.6861582055762283E-11 & 0.3862018590575813E-10 & 0.3101695335777539E+00
\\\hline
\end{tabular}}\end{center}                                                         

\begin{center}{\scriptsize\begin{tabular}{|c|c|c|c|}
\hline
     ${-++++++}$     & $\epsilon^{-2}$        & $\epsilon^{-1}$        & $\epsilon^{0}$
\\\hline
   tree            &                        &                        & 0.1648597081617964E-14
\\ tree \GZ        &                        &                        & 0.0000000000000000E+00
\\ unit \GZ        & 0.3678212874319657E-12 & 0.7209572152581734E-12 & 0.1920528148108100E+00
\\ anly \GZ        & 0.2713533399763100E-14 & 0.8924875144594874E-14 & 0.1920528147653950E+00
\\ tnsr            & 0.4384180396142481E-10 & 0.2104965953958677E-09 & 0.1920528150979991E+00
\\\hline
\end{tabular}}\end{center}                                                         

\begin{center}{\scriptsize\begin{tabular}{|c|c|c|c|}
\hline
     ${--+++++}$     & $\epsilon^{-2}$        & $\epsilon^{-1}$        & $\epsilon^{0}$
\\\hline
   tree            &                        &                        & 0.2106612834594487E+01
\\ tree \GZ        &                        &                        & 0.2106612834594490E+01
\\ unit \GZ        & 0.1474628984216140E+02 & 0.4850089396312140E+02 & 0.8731521551387900E+02
\\ anly \GZ        & 0.1474628984216140E+02 & 0.4850089396312130E+02 & 0.8731521551386510E+02
\\ tnsr            & 0.1474628984215924E+02 & 0.4850089396308871E+02 & 0.8731521551379141E+02
\\\hline
\end{tabular}}\end{center}                                                         

\begin{center}{\scriptsize\begin{tabular}{|c|c|c|c|}
\hline
     ${-+-+-+-}$     & $\epsilon^{-2}$        & $\epsilon^{-1}$        & $\epsilon^{0}$
\\\hline
   tree            &                        &                        & 0.1101865680944418E+00
\\ tree \GZ        &                        &                        & 0.1101865680944420E+00
\\ unit \GZ        & 0.7713059766610930E+00 & 0.2536843489960730E+01 & 0.5933610502945470E+01
\\ anly \GZ        & 0.7713059766610950E+00 & 0.2536843489960750E+01 &                       
\\ tnsr            & 0.7713059766434806E+00 & 0.2536843489868709E+01 & 0.5933610502629016E+01
\\\hline
\end{tabular}}\end{center}                                                         

\begin{center}{\scriptsize\begin{tabular}{|c|c|c|c|}
\hline
     ${+-+-+-+}$     & $\epsilon^{-2}$        & $\epsilon^{-1}$        & $\epsilon^{0}$
\\\hline
   tree            &                        &                        & 0.1101865680944424E+00
\\ tree \GZ        &                        &                        & 0.1101865680944420E+00
\\ unit \GZ        & 0.7713059766610930E+00 & 0.2536843489960740E+01 & 0.6042012409916140E+01
\\ anly \GZ        & 0.7713059766610950E+00 & 0.2536843489960750E+01 &                       
\\ tnsr            & 0.7713059766436269E+00 & 0.2536843489862686E+01 & 0.6042012409615074E+01
\\\hline
\end{tabular}}\end{center}                                                         

\subsection{$\mult=8$}

\begin{center}{\scriptsize\begin{tabular}{|c|c|c|c|}
\hline
     ${++++++++}$    & $\epsilon^{-2}$        & $\epsilon^{-1}$        & $\epsilon^{0}$
\\\hline
   tree            &                        &                        & 0.2961844168510185E-15
\\ tree \GZ        &                        &                        & 0.0000000000000000E+00
\\ unit \GZ        & 0.0000000000000000E+00 & 0.0000000000000000E+00 & 0.1967006006956910E+00
\\ anly \GZ        & 0.3853462894343397E-14 & 0.1441159379540454E-13 & 0.1967006007382010E+00
\\ tnsr            & 0.1336896918161384E-10 & 0.5652403097216611E-10 & 0.1967006012222209E+00
\\\hline
\end{tabular}}\end{center}                                                         

\begin{center}{\scriptsize\begin{tabular}{|c|c|c|c|}
\hline
     ${-+++++++}$    & $\epsilon^{-2}$        & $\epsilon^{-1}$        & $\epsilon^{0}$
\\\hline
   tree            &                        &                        & 0.2951926495253574E-14
\\ tree \GZ        &                        &                        & 0.2257277386254959E-14
\\ unit \GZ        & 0.0000000000000000E+00 & 0.1965638104048654E-09 & 0.5287747164930630E+00
\\ anly \GZ        & 0.1805821909003967E-13 & 0.6753606439965886E-13 & 0.5287747176521700E+00
\\ tnsr            & 0.2640076774290492E-10 & 0.1008615996797149E-09 & 0.5287747173953582E+00
\\\hline
\end{tabular}}\end{center}                                                         

\begin{center}{\scriptsize\begin{tabular}{|c|c|c|c|}
\hline
     ${--++++++}$    & $\epsilon^{-2}$        & $\epsilon^{-1}$        & $\epsilon^{0}$
\\\hline
   tree            &                        &                        & 0.4333189194669586E+01
\\ tree \GZ        &                        &                        & 0.4333189194669600E+01
\\ unit \GZ        & 0.3466551355735610E+02 & 0.1296458053471450E+03 & 0.2742997734349260E+03
\\ anly \GZ        & 0.3466551355735680E+02 & 0.1296458052914090E+03 & 0.2742997734349000E+03
\\ tnsr            & 0.3466551355735630E+02 & 0.1296458052913800E+03 & 0.2742997734347139E+03
\\\hline
\end{tabular}}\end{center}                                                         

\begin{center}{\scriptsize\begin{tabular}{|c|c|c|c|}
\hline
     ${-+-+-+-+}$    & $\epsilon^{-2}$        & $\epsilon^{-1}$        & $\epsilon^{0}$
\\\hline
   tree            &                        &                        & 0.7261522613885360E-01
\\ tree \GZ        &                        &                        & 0.7261522613885579E-01
\\ unit \GZ        & 0.5809218091107730E+00 & 0.2172593680235970E+01 & 0.5476303819766790E+01
\\ anly \GZ        & 0.5809218091108460E+00 & 0.2172593682447690E+01 &                       
\\ tnsr            & 0.5809218091244186E+00 & 0.2172593682447645E+01 & 0.5476303819972849E+01
\\\hline
\end{tabular}}\end{center}                                                         

\begin{center}{\scriptsize\begin{tabular}{|c|c|c|c|}
\hline
     ${+-+-+-+-}$    & $\epsilon^{-2}$        & $\epsilon^{-1}$        & $\epsilon^{0}$
\\\hline
   tree            &                        &                        & 0.7261522613885418E-01
\\ tree \GZ        &                        &                        & 0.7261522613885579E-01
\\ unit \GZ        & 0.5809218091108620E+00 & 0.2172593687810420E+01 & 0.4925500546307290E+01
\\ anly \GZ        & 0.5809218091108460E+00 & 0.2172593682447690E+01 &                       
\\ tnsr            & 0.5809218091153071E+00 & 0.2172593682480466E+01 & 0.4925500546470287E+01
\\\hline
\end{tabular}}\end{center}                                                         

\subsection{$\mult=9$}

\begin{center}{\scriptsize\begin{tabular}{|c|c|c|c|}
\hline
     ${+++++++++}$   & $\epsilon^{-2}$        & $\epsilon^{-1}$        & $\epsilon^{0}$
\\\hline
   tree            &                        &                        & 0.3090869336705567E-13
\\ tree \GZ        &                        &                        & 0.2992915640032351E-13
\\ unit \GZ        & 0.4860269836292316E-11 & 0.1845193695700690E-07 & 0.5666555617062950E+01
\\ anly \GZ        & 0.2693624076029116E-12 & 0.1176695244346755E-11 & 0.5666555580473110E+01
\\ tnsr            & 0.1242803468718531E-06 & 0.1395624387049265E-05 & 0.5666561857344509E+01
\\\hline
\end{tabular}}\end{center}                                                         

\begin{center}{\scriptsize\begin{tabular}{|c|c|c|c|}
\hline
     ${-++++++++}$   & $\epsilon^{-2}$        & $\epsilon^{-1}$        & $\epsilon^{0}$
\\\hline
   tree            &                        &                        & 0.7191085595712448E-13
\\ tree \GZ        &                        &                        & 0.9114087930248735E-13
\\ unit \GZ        & 0.3938371378126140E-10 & 0.2340429860576292E-07 & 0.1062086460614280E+01
\\ anly \GZ        & 0.8202679137223861E-12 & 0.3583296428617654E-11 & 0.1062086467981750E+01
\\ tnsr            & 0.1761715389271764E-05 & 0.2085999886200371E-04 & 0.1062166559796595E+01
\\\hline
\end{tabular}}\end{center}                                                         

\begin{center}{\scriptsize\begin{tabular}{|c|c|c|c|}
\hline
     ${--+++++++}$   & $\epsilon^{-2}$        & $\epsilon^{-1}$        & $\epsilon^{0}$
\\\hline
   tree            &                        &                        & 0.3232296679455183E+02
\\ tree \GZ        &                        &                        & 0.3232296679455080E+02
\\ unit \GZ        & 0.2909067010969220E+03 & 0.1270810334861320E+04 & 0.3625430616705210E+04
\\ anly \GZ        & 0.2909067011509570E+03 & 0.1270810336301850E+04 & 0.3625430616705940E+04
\\ tnsr            & 0.2909066975088621E+03 & 0.1270810392170893E+04 & 0.3625430209440870E+04
\\\hline
\end{tabular}}\end{center}                                                         

\begin{center}{\scriptsize\begin{tabular}{|c|c|c|c|}
\hline
     ${-+-+-+-+-}$   & $\epsilon^{-2}$        & $\epsilon^{-1}$        & $\epsilon^{0}$
\\\hline
   tree            &                        &                        & 0.4535219663678330E+00
\\ tree \GZ        &                        &                        & 0.4535219663679500E+00
\\ unit \GZ        & 0.4081697696661860E+01 & 0.1783067767208140E+02 & 0.5710639504628740E+02
\\ anly \GZ        & 0.4081697697311550E+01 & 0.1783067768078440E+02 &                       
\\ tnsr            & 0.4081697504765963E+01 & 0.1783065940829036E+02 & 0.5710625504238800E+02
\\\hline
\end{tabular}}\end{center}                                                         

\begin{center}{\scriptsize\begin{tabular}{|c|c|c|c|}
\hline
     ${+-+-+-+-+}$   & $\epsilon^{-2}$        & $\epsilon^{-1}$        & $\epsilon^{0}$
\\\hline
   tree            &                        &                        & 0.4535219663678577E+00
\\ tree \GZ        &                        &                        & 0.4535219663679500E+00
\\ unit \GZ        & 0.4081697696620550E+01 & 0.1783067764548420E+02 & 0.5501538077075760E+02
\\ anly \GZ        & 0.4081697697311550E+01 & 0.1783067768078440E+02 &                       
\\ tnsr            & 0.4081697558409711E+01 & 0.1783066268732069E+02 & 0.5501537931596933E+02
\\\hline
\end{tabular}}\end{center}                                                         

\subsection{$\mult=10$}

\begin{center}{\scriptsize\begin{tabular}{|c|c|c|c|}
\hline
     ${++++++++++}$  & $\epsilon^{-2}$        & $\epsilon^{-1}$        & $\epsilon^{0}$
\\\hline
   tree            &                        &                        & 0.8422572777655544E-13
\\ tree \GZ        &                        &                        & 0.7645214091184737E-13
\\ unit \GZ        & 0.2616999209810146E-12 & 0.7453142378465002E-06 & 0.1843490112846700E+02
\\ anly \GZ        & 0.7645214091184737E-12 & 0.3853184186191476E-11 & 0.1843490112846710E+02
\\ tnsr            & 0.1199729722684045E-06 & 0.1810959673219180E-05 & 0.1843487909054984E+02
\\\hline
\end{tabular}}\end{center}                                                         

\begin{center}{\scriptsize\begin{tabular}{|c|c|c|c|}
\hline
     ${-+++++++++}$  & $\epsilon^{-2}$        & $\epsilon^{-1}$        & $\epsilon^{0}$
\\\hline
   tree            &                        &                        & 0.1538190662118770E-12
\\ tree \GZ        &                        &                        & 0.3138928592085274E-12
\\ unit \GZ        & 0.1729567134060808E-10 & 0.3462486730362966E-05 & 0.1411806902836740E+02
\\ anly \GZ        & 0.3138928592085274E-11 & 0.1582018484813023E-10 & 0.1411806902836920E+02
\\ tnsr            & 0.9039805077823879E-06 & 0.8416576413616293E-05 & 0.1411799961139769E+02
\\\hline
\end{tabular}}\end{center}                                                         

\begin{center}{\scriptsize\begin{tabular}{|c|c|c|c|}
\hline
     ${--++++++++}$  & $\epsilon^{-2}$        & $\epsilon^{-1}$        & $\epsilon^{0}$
\\\hline
   tree            &                        &                        & 0.4899726956663458E+03
\\ tree \GZ        &                        &                        & 0.4899726956663410E+03
\\ unit \GZ        & 0.4899726956656070E+04 & 0.2469460004000990E+05 & 0.7584491014580890E+05
\\ anly \GZ        & 0.4899726956663410E+04 & 0.2469460004768270E+05 & 0.7584491014578140E+05
\\ tnsr            & 0.4899726958888808E+04 & 0.2469460005194075E+05 & 0.7584491017379209E+05
\\\hline
\end{tabular}}\end{center}                                                         

\begin{center}{\scriptsize\begin{tabular}{|c|c|c|c|}
\hline
     ${-+-+-+-+-+}$  & $\epsilon^{-2}$        & $\epsilon^{-1}$        & $\epsilon^{0}$
\\\hline
   tree            &                        &                        & 0.9346113720088734E+01
\\ tree \GZ        &                        &                        & 0.9346113720089020E+01
\\ unit \GZ        & 0.9346113719987591E+02 & 0.4710436787027110E+03 & 0.1481274476056640E+04
\\ anly \GZ        & 0.9346113720089021E+02 & 0.4710436772479390E+03 &                       
\\ tnsr            & 0.9346113676127024E+02 & 0.4710436797756468E+03 & 0.1481274475543962E+04
\\\hline
\end{tabular}}\end{center}                                                         

\begin{center}{\scriptsize\begin{tabular}{|c|c|c|c|}
\hline
     ${+-+-+-+-+-}$  & $\epsilon^{-2}$        & $\epsilon^{-1}$        & $\epsilon^{0}$
\\\hline
   tree            &                        &                        & 0.9346113720088464E+01
\\ tree \GZ        &                        &                        & 0.9346113720089020E+01
\\ unit \GZ        & 0.9346113719956180E+02 & 0.4710436740057420E+03 & 0.1503970258031110E+04
\\ anly \GZ        & 0.9346113720089021E+02 & 0.4710436772479390E+03 &                       
\\ tnsr            & 0.9346113696575948E+02 & 0.4710436794794809E+03 & 0.1503970260993868E+04
\\\hline
\end{tabular}}\end{center}

%% file: appendixB.tex
This appendix contains results with one massless quark-loop exclusively without the gluoninc loops, and the full one-loop amplitude including one massless quark-loop.
The former are labelled with ``excl'', and the latter with ``incl''.
The phase-space points are the same as in \Appendix{App1}, and are given in \cite{Giele:2008bc}.
Presented is the absolute value of the coefficients of $\epsilon^{-2}$, $\epsilon^{-1}$ and $\epsilon^{0}$ of the amplitudes.
Results labelled with ``tree'' give the latter value for the tree-level amplitude.
The upper-left box in each table gives the helicity configuration.

\subsection{$\mult=6$}

\begin{center}{\scriptsize\begin{tabular}{|c|c|c|c|}
\hline
       ${++++++}$    & $\epsilon^{-2}$        & $\epsilon^{-1}$        & $\epsilon^{0}$
\\\hline
   tree            &                        &                        & 0.2215923815877299E-14
\\ excl            & 0.3444649545719067E-11 & 0.3563314196395845E-11 & 0.1766021612412397E+00
\\ incl            & 0.5484082406081639E-11 & 0.5497590035137738E-10 & 0.3532043223779056E+00
\\\hline
\end{tabular}}\end{center}                                                         

\begin{center}{\scriptsize\begin{tabular}{|c|c|c|c|}
\hline
     ${-+++++}$      & $\epsilon^{-2}$        & $\epsilon^{-1}$        & $\epsilon^{0}$
\\\hline
   tree            &                        &                        & 0.1947124136075826E-13
\\ excl            & 0.3588818407048125E-10 & 0.2958732120038221E-09 & 0.1086655684552452E+01
\\ incl            & 0.8158777253122160E-10 & 0.1238241794955264E-08 & 0.2173311369421295E+01
\\\hline
\end{tabular}}\end{center}                                                         

\begin{center}{\scriptsize\begin{tabular}{|c|c|c|c|}
\hline
     ${--++++}$      & $\epsilon^{-2}$        & $\epsilon^{-1}$        & $\epsilon^{0}$
\\\hline
   tree            &                        &                        & 0.2849128165044324E+02
\\ excl            & 0.3083625492388671E-10 & 0.6331395922312280E+01 & 0.2228503047232156E+02
\\ incl            & 0.1709476899026306E+03 & 0.6094559286552980E+03 & 0.1363324204445743E+04
\\\hline
\end{tabular}}\end{center}                                                         

\begin{center}{\scriptsize\begin{tabular}{|c|c|c|c|}
\hline
     ${-+-+-+}$      & $\epsilon^{-2}$        & $\epsilon^{-1}$        & $\epsilon^{0}$
\\\hline
   tree            &                        &                        & 0.3138715395008085E+01
\\ excl            & 0.8282931375686374E-10 & 0.6974923101314857E+00 & 0.3642128932022815E+01
\\ incl            & 0.1883229236997337E+02 & 0.6714014235334223E+02 & 0.1496072786704808E+03
\\\hline
\end{tabular}}\end{center}                                                         

\begin{center}{\scriptsize\begin{tabular}{|c|c|c|c|}
\hline
     ${+-+-+-}$      & $\epsilon^{-2}$        & $\epsilon^{-1}$        & $\epsilon^{0}$
\\\hline
   tree            &                        &                        & 0.3138715395008091E+01
\\ excl            & 0.1130982098649374E-10 & 0.6974923098246018E+00 & 0.1399955550457513E+01
\\ incl            & 0.1883229236989885E+02 & 0.6714014235419050E+02 & 0.1526894583161683E+03
\\\hline
\end{tabular}}\end{center}                                                         

\subsection{$\mult=7$}

\begin{center}{\scriptsize\begin{tabular}{|c|c|c|c|}
\hline
     ${+++++++}$     & $\epsilon^{-2}$        & $\epsilon^{-1}$        & $\epsilon^{0}$
\\\hline
   tree            &                        &                        & 0.2703922191931054E-14
\\ excl            & 0.2734354818049004E-12 & 0.1535659632517864E-11 & 0.1033898444969325E+00
\\ incl            & 0.5758778544760509E-12 & 0.3391282384997409E-11 & 0.2067796889938174E+00
\\\hline
\end{tabular}}\end{center}                                                         

\begin{center}{\scriptsize\begin{tabular}{|c|c|c|c|}
\hline
     ${-++++++}$     & $\epsilon^{-2}$        & $\epsilon^{-1}$        & $\epsilon^{0}$
\\\hline
   tree            &                        &                        & 0.1648597081617964E-14
\\ excl            & 0.7829286354785003E-11 & 0.4038378546519624E-10 & 0.6401760500069484E-01
\\ incl            & 0.1149200607547395E-10 & 0.6152252106547420E-10 & 0.1280352099351288E+00
\\\hline
\end{tabular}}\end{center}                                                         

\begin{center}{\scriptsize\begin{tabular}{|c|c|c|c|}
\hline
     ${--+++++}$     & $\epsilon^{-2}$        & $\epsilon^{-1}$        & $\epsilon^{0}$
\\\hline
   tree            &                        &                        & 0.2106612834594487E+01
\\ excl            & 0.3039208000395293E-12 & 0.4681361854653245E+00 & 0.7724537168857577E+00
\\ incl            & 0.1474628984215543E+02 & 0.4815969760605956E+02 & 0.8723793738684391E+02
\\\hline
\end{tabular}}\end{center}                                                         

\begin{center}{\scriptsize\begin{tabular}{|c|c|c|c|}
\hline
     ${-+-+-+-}$     & $\epsilon^{-2}$        & $\epsilon^{-1}$        & $\epsilon^{0}$
\\\hline
   tree            &                        &                        & 0.1101865680944418E+00
\\ excl            & 0.4415997997437337E-11 & 0.2448590404383259E-01 & 0.8801500600694001E-01
\\ incl            & 0.7713059766680262E+00 & 0.2518997184786409E+01 & 0.5849371002103936E+01
\\\hline
\end{tabular}}\end{center}                                                         

\begin{center}{\scriptsize\begin{tabular}{|c|c|c|c|}
\hline
     ${+-+-+-+}$     & $\epsilon^{-2}$        & $\epsilon^{-1}$        & $\epsilon^{0}$
\\\hline
   tree            &                        &                        & 0.1101865680944424E+00
\\ excl            & 0.4278136193897277E-11 & 0.2448590401493046E-01 & 0.9578987033471566E-01
\\ incl            & 0.7713059766681208E+00 & 0.2518997184763565E+01 & 0.5952686996619508E+01
\\\hline
\end{tabular}}\end{center}                                                         

\subsection{$\mult=8$}

\begin{center}{\scriptsize\begin{tabular}{|c|c|c|c|}
\hline
     ${++++++++}$    & $\epsilon^{-2}$        & $\epsilon^{-1}$        & $\epsilon^{0}$
\\\hline
   tree            &                        &                        & 0.2961844168510185E-15
\\ excl            & 0.6535721923323608E-12 & 0.2946477554703654E-10 & 0.6556686705957517E-01
\\ incl            & 0.6936524320241993E-11 & 0.2689757765745469E-10 & 0.1311337340887097E+00
\\\hline
\end{tabular}}\end{center}                                                         

\begin{center}{\scriptsize\begin{tabular}{|c|c|c|c|}
\hline
     ${-+++++++}$    & $\epsilon^{-2}$        & $\epsilon^{-1}$        & $\epsilon^{0}$
\\\hline
   tree            &                        &                        & 0.2951926495253574E-14
\\ excl            & 0.2993517822283384E-11 & 0.4138014710832325E-10 & 0.1762582391337885E+00
\\ incl            & 0.1682824582337319E-10 & 0.2282288551418147E-10 & 0.3525164784441872E+00
\\\hline
\end{tabular}}\end{center}                                                         

\begin{center}{\scriptsize\begin{tabular}{|c|c|c|c|}
\hline
     ${--++++++}$    & $\epsilon^{-2}$        & $\epsilon^{-1}$        & $\epsilon^{0}$
\\\hline
   tree            &                        &                        & 0.4333189194669586E+01
\\ excl            & 0.2048884758899206E-11 & 0.9629309321605227E+00 & 0.1952623627599051E+01
\\ incl            & 0.3466551355734642E+02 & 0.1288994364902356E+03 & 0.2738716469885952E+03
\\\hline
\end{tabular}}\end{center}                                                         

\begin{center}{\scriptsize\begin{tabular}{|c|c|c|c|}
\hline
     ${-+-+-+-+}$    & $\epsilon^{-2}$        & $\epsilon^{-1}$        & $\epsilon^{0}$
\\\hline
   tree            &                        &                        & 0.7261522613885360E-01
\\ excl            & 0.6007828400852260E-11 & 0.1613671694024866E-01 & 0.7432198837943639E-01
\\ incl            & 0.5809218091173581E+00 & 0.2160086095767551E+01 & 0.5416218143354230E+01
\\\hline
\end{tabular}}\end{center}                                                         

\begin{center}{\scriptsize\begin{tabular}{|c|c|c|c|}
\hline
     ${+-+-+-+-}$    & $\epsilon^{-2}$        & $\epsilon^{-1}$        & $\epsilon^{0}$
\\\hline
   tree            &                        &                        & 0.7261522613885418E-01
\\ excl            & 0.2176051463652288E-11 & 0.1613671696641898E-01 & 0.3887448345810546E-01
\\ incl            & 0.5809218091128370E+00 & 0.2160086095780097E+01 & 0.4897808762927628E+01
\\\hline
\end{tabular}}\end{center}                                                         

\subsection{$\mult=9$}

\begin{center}{\scriptsize\begin{tabular}{|c|c|c|c|}
\hline
     ${+++++++++}$   & $\epsilon^{-2}$        & $\epsilon^{-1}$        & $\epsilon^{0}$
\\\hline
   tree            &                        &                        & 0.3090869336705567E-13
\\ excl            & 0.1679041191381272E-07 & 0.4256557696335983E-06 & 0.1888851178320152E+01
\\ incl            & 0.9845976672993657E-07 & 0.1169508469550796E-05 & 0.3777703015967018E+01
\\\hline
\end{tabular}}\end{center}                                                         

\begin{center}{\scriptsize\begin{tabular}{|c|c|c|c|}
\hline
     ${-++++++++}$   & $\epsilon^{-2}$        & $\epsilon^{-1}$        & $\epsilon^{0}$
\\\hline
   tree            &                        &                        & 0.7191085595712448E-13
\\ excl            & 0.7127945940983221E-06 & 0.7142546418827421E-05 & 0.3540723883196215E+00
\\ incl            & 0.1124121726807117E-05 & 0.1415224198572896E-04 & 0.7080938879221086E+00
\\\hline
\end{tabular}}\end{center}                                                         

\begin{center}{\scriptsize\begin{tabular}{|c|c|c|c|}
\hline
     ${--+++++++}$   & $\epsilon^{-2}$        & $\epsilon^{-1}$        & $\epsilon^{0}$
\\\hline
   tree            &                        &                        & 0.3232296679455183E+02
\\ excl            & 0.3697111686288034E-05 & 0.7182873316896260E+01 & 0.8760538757567446E+01
\\ incl            & 0.2909066988869654E+03 & 0.1264862610175780E+04 & 0.3621189040116724E+04
\\\hline
\end{tabular}}\end{center}                                                         

\begin{center}{\scriptsize\begin{tabular}{|c|c|c|c|}
\hline
     ${-+-+-+-+-}$   & $\epsilon^{-2}$        & $\epsilon^{-1}$        & $\epsilon^{0}$
\\\hline
   tree            &                        &                        & 0.4535219663678330E+00
\\ excl            & 0.2296878137953301E-06 & 0.1007747715318806E+00 & 0.3472201563087314E+00
\\ incl            & 0.4081697557451384E+01 & 0.1774721263709833E+02 & 0.5675986177672981E+02
\\\hline
\end{tabular}}\end{center}                                                         

\begin{center}{\scriptsize\begin{tabular}{|c|c|c|c|}
\hline
     ${+-+-+-+-+}$   & $\epsilon^{-2}$        & $\epsilon^{-1}$        & $\epsilon^{0}$
\\\hline
   tree            &                        &                        & 0.4535219663678577E+00
\\ excl            & 0.2530328589867088E-06 & 0.1007792647137644E+00 & 0.3151037218153409E+00
\\ incl            & 0.4081697604003098E+01 & 0.1774721648717686E+02 & 0.5470112910999667E+02
\\\hline
\end{tabular}}\end{center}                                                         

\subsection{$\mult=10$}

\begin{center}{\scriptsize\begin{tabular}{|c|c|c|c|}
\hline
     ${++++++++++}$  & $\epsilon^{-2}$        & $\epsilon^{-1}$        & $\epsilon^{0}$
\\\hline
   tree            &                        &                        & 0.8422572777655544E-13
\\ excl            & 0.7742516776772886E-07 & 0.8116698643478135E-06 & 0.6144957172059176E+01
\\ incl            & 0.1833646911715754E-06 & 0.2136772265813164E-05 & 0.1228991105613530E+02
\\\hline
\end{tabular}}\end{center}                                                         

\begin{center}{\scriptsize\begin{tabular}{|c|c|c|c|}
\hline
     ${-+++++++++}$  & $\epsilon^{-2}$        & $\epsilon^{-1}$        & $\epsilon^{0}$
\\\hline
   tree            &                        &                        & 0.1538190662118770E-12
\\ excl            & 0.5628204459080735E-06 & 0.3925867464691803E-05 & 0.4705985971256853E+01
\\ incl            & 0.1437262034778626E-05 & 0.8702188415939938E-05 & 0.9412025021303032E+01
\\\hline
\end{tabular}}\end{center}                                                         

\begin{center}{\scriptsize\begin{tabular}{|c|c|c|c|}
\hline
     ${--++++++++}$  & $\epsilon^{-2}$        & $\epsilon^{-1}$        & $\epsilon^{0}$
\\\hline
   tree            &                        &                        & 0.4899726956663458E+03
\\ excl            & 0.1254733694226626E-05 & 0.1088828222285149E+03 & 0.2631225655494603E+03
\\ incl            & 0.4899726955335932E+04 & 0.2460028096277122E+05 & 0.7571494684904416E+05
\\\hline
\end{tabular}}\end{center}                                                         

\begin{center}{\scriptsize\begin{tabular}{|c|c|c|c|}
\hline
     ${-+-+-+-+-+}$  & $\epsilon^{-2}$        & $\epsilon^{-1}$        & $\epsilon^{0}$
\\\hline
   tree            &                        &                        & 0.9346113720088734E+01
\\ excl            & 0.4577998694131886E-06 & 0.2076909542983419E+01 & 0.7503126329349836E+01
\\ incl            & 0.9346113693450476E+02 & 0.4692445627736100E+03 & 0.1474914895661399E+04
\\\hline
\end{tabular}}\end{center}                                                         

\begin{center}{\scriptsize\begin{tabular}{|c|c|c|c|}
\hline
     ${+-+-+-+-+-}$  & $\epsilon^{-2}$        & $\epsilon^{-1}$        & $\epsilon^{0}$
\\\hline
   tree            &                        &                        & 0.9346113720088464E+01
\\ excl            & 0.4345102202108285E-06 & 0.2076912724815172E+01 & 0.6824080572856147E+01
\\ incl            & 0.9346113705011550E+02 & 0.4692445616703819E+03 & 0.1498766884054935E+04
\\\hline
\end{tabular}}\end{center}